\documentclass[10pt,english,english,nofootinbib,notitlepage,superscriptaddress,aps,prd]{revtex4-1}
\usepackage{enumerate}
\usepackage{amsmath}
\usepackage{amsfonts}
\usepackage{amssymb}
\usepackage[utf8]{inputenc}
\usepackage[T1]{fontenc}
\usepackage{mathtools}
\usepackage{wasysym}
\usepackage{accents}
\usepackage[svgnames]{xcolor}
\usepackage[colorlinks=true, pdfstartview=FitV, linkcolor=blue, citecolor=red, urlcolor=magenta]{hyperref}
\usepackage{url}
\usepackage{graphicx}
\usepackage{subfig}
\usepackage{bm}
\usepackage{tikz}
\usetikzlibrary{arrows,decorations.markings}
\usepackage{csquotes}

\usepackage{esint}
\usepackage{color}
\DeclareMathOperator{\arccosh}{arccosh}
\DeclareMathOperator{\arcsinh}{arcsinh}

\begin{document}

\title{Anti-de Sitter Momentum Space in 3D and 4D Quantum Gravity}

\author{Giovanni Amelino-Camelia}
\email{amelino@na.infn.it}
\affiliation{Dipartimento di Fisica Ettore Pancini, Universit\`a 
di Napoli ``Federico II'', and INFN, Sezione di Napoli,\\ Complesso Univ. Monte S. Angelo, I-80126 Napoli, Italy}

\author{Iarley P. Lobo}
\email{lobofisica@gmail.com}
\affiliation{Dipartimento di Fisica, Universit\`a di Roma ``La Sapienza\textquotedbl{},
P.le A. Moro 2, 00185 Roma, Italy}

\affiliation{Department of Chemistry and Physics, Federal University of Para\'iba, Rodovia BR 079 - km 12, 58397-000 Areia-PB,  Brazil}

\affiliation{Physics Department, Federal University of Lavras, Caixa Postal 3037, 37200-900 Lavras-MG, Brazil}

\author{Giovanni Palmisano}
\email{giovanni.palmisano1@gmail.com}
\affiliation{Dipartimento di Fisica, Universit\`a di Roma ``La Sapienza\textquotedbl{},
P.le A. Moro 2, 00185 Roma, Italy}
\affiliation{INFN, Sez.~Roma1, P.le A. Moro 2, 00185 Roma, Italy}

\date{\today}
\begin{abstract}
There has been strong interest in the possibility that in the quantum-gravity realm 
momentum space might be curved, mainly focusing, especially for what concerns
phenomenological implications, on the case of a de Sitter momentum space.
We here take as starting point the known fact that 
quantum gravity coupled to matter in $2+1$ spacetime dimensions
gives rise to an effective picture characterized by a momentum space
with anti-de Sitter geometry, and we point out some key properties of
$2+1$-dimensional anti-de Sitter momentum space. We observe that it is impossible to implement all of these properties in theories with a $3+1$-dimensional anti-de Sitter momentum space, and we then investigate, with the aim of providing guidance to the relevant phenomenology focusing on possible modified laws of conservation of momenta, the implications of  giving up, in the $3+1$-dimensional case, some of the properties of the $2+1$-dimensional case.
 \end{abstract}

\maketitle


\section{Introduction}
Over the last few years a growing number of quantum-gravity studies has focused
on the possibility of having a Planck-scale-curved momentum space, with the associated emergence
of DSR-relativistic \cite{Amelinodsr1} spacetime symmetries (``DSR'' stands for
 ``doubly-special'', or occasionally ``deformed-special'', relativity).
 This concerns the
 possibility that in the quantum-gravity realm relativistic symmetries might be deformed in such a way that one would not only have a speed scale $c$ as non-trivial relativistic invariant, but also a length/inverse-momentum scale.

 Besides its intrinsic conceptual interest, this scenario has also been studied as a rare example of quantum-gravity scenario which provides some viable phenomenological avenues \cite{Amelinodsr2,Amelino-Camelia2013,Addazi:2021xuf}.
Additional elements of motivation for such studies are provided
by the fact that $2+1$-dimensional quantum gravity, the only quantum-gravity
we have managed to solve \cite{freidellivineeffective,MatschullWelling},
 does indeed lead to DSR-relativistic symmetries and curvature of momentum space.
In $2+1$ spacetime dimensions the gravitational
degrees of freedom are topological and many results can be obtained
more easily then in the $3+1$-dimensional case. In particular, by
describing particles as  topological defects (``punctures'') of the spacetime
manifold, it is possible to integrate out the gravitational degrees
of freedom and the resulting effective model is characterized \cite{freidellivineeffective}
by a curved momentum space
within a DSR-relativistic picture \cite{RellocinAdS}.

The 4D (3+1-dimensional) quantum gravity we care about surely is very different
from 3D quantum gravity, but, since the two theories have in common
some of their aspects of conceptual complexity, it may well be the case that they also have
some features in common. Perhaps the most appealing argument for
the emergence of  DSR-relativistic symmetries in the 3+1D context
 is found
 in some recent
 studies \cite{Amelino-Camelia:2016gfx,Bojowald:2012ux,Brahma:2016tsq,Cianfrani:2016ogm},
  where models inspired by the  loop-quantum-gravity approach
were shown to lead to modifications of the hypersurface deformation algebra:
in turn those deformations of the hypersurface deformation algebra were
found to give rise to DSR-relativistic symmetries in
the appropriate limiting regime \cite{Amelino-Camelia:2016gfx,Bojowald:2012ux}.
It is however still difficult to guess from these preliminary 4D results and
from what is known about 3D quantum gravity what is
the type of DSR-relativistic symmetries that should be favoured for 4D quantum gravity. Most related studies assumed a de Sitter geometry of momentum space
 (see, {\it e.g.}, Refs.\cite{GiuliaFlavio,Carmona:2019fwf} and references therein). To
 our knowledge the only study considering  the anti-de Sitter momentum-space possibility
 is the one reported in Ref.\cite{Arzano:2014jua}, but it focused exclusively on issues relevant
 for single-particle phenomenology (mainly correction terms for the laws of propagation and
 a mechanism for dimensional reduction in the ultra-violet regime).

We find it to be particularly
noteworthy that the DSR-relativistic symmetries of
 3D quantum gravity are codified in a momentum space with anti-de Sitter geometry \cite{freidellivineeffective,MatschullWelling},
 in a way that is fully compatible with the so-called ``Relative Locality framework''
 \cite{RellocinAdS,principle}, and admits formulation in terms of
 a Hopf algebra of relativistic symmetries \cite{Schroers2011}. We here here take as working assumption that also in 4D quantum gravity it should be
possible to describe curvature of momentum space within the Relative Locality framework,
and we investigate whether or not one gets
a sensible picture by adopting also for 4D quantum gravity a momentum space
with anti-de Sitter geometry. In fact, quantum gravity effects being manifest in the form of curved momentum space and Relative Locality has been recently explored in different frameworks \cite{Freidel:2013zga,Freidel:2015pka,Freidel:2016pls,Freidel:2018apz,Freidel:2022ryr,Berglund:2023vrm}. Of particular significance for phenomenology are our
results concerning possible laws of composition of momenta on an  anti-de Sitter momentum space,
since they can affect particle reactions (see, {\it e.g.}, Refs.\cite{Carmona:2014lqa} and references therein).

A key aspect of our analysis concerns the identification of some properties that
characterize the 3D law of composition of momenta, reflecting properties
of the anti-de Sitter geometry of momentum space: some interesting implications arise
investigating whether the same
properties can be imposed in the 4D setting.
Empowered by the geometric tools introduced by the relative-locality framework \cite{principle},
we find appropriate to focus on four properties of the law of composition
of momenta, which we here list but will be
more carefully discussed later in the manuscript:

\noindent
$R1$) Existence of identity;

\noindent
$R2$)
Associativity;

\noindent
$R3$) Isometric action of the composition law on the
geodesic distance in momentum space;

\noindent
 $R4$) Relativistic covariance
of the theory.

It is noteworthy that these 4 properties
not only are present in 3D quantum gravity,
but are also present in a particular formulation of de Sitter momentum space,
the one amenable to interpretation  \cite{GiuliaFlavio,arzano2011anatomy} in terms
of the $\kappa$-Poincar\'e
Hopf algebra \cite{lukierski1991q,Lukierski1992344,MajidRUEGG,Amelinodsr1}.
We here conjecture that the availability of a Hopf-algebra interpretation
is connected to the validity of the properties $R1$, $R2$, $R3$ and $R4$. 

Section \ref{sec:Relative-Locality-Analysis} is devoted to a summary of relevant 3D-quantum-gravity results
and results on the relative-locality momentum space which admits $\kappa$-Poincar\'e
interpretation.
In section \ref{sec:-Dimensional-anti-de} we
show that adopting an anti-de Sitter momentum space for 4D quantum gravity
one is unable to enforce the requirements $R1$,
$R2$, $R3$ and $R4$: essentially there is 
an incompatibility between $R3$ and $R4$
for 4D anti-de Sitter momentum space (as implemented in the relative-locality framework).
This is consistent with the findings of previous studies (see, in particular, the appendix of Ref.\cite{Carmona:2019fwf}), which however focused on scenarios with spatial isotropy, whereas we here extend the result to the anisotropic case.
Our main objective is to provide guidance to the relevant phenomenology:
having found that
$R1$,
$R2$, $R3$ and $R4$ cannot be simultaneously implemented within
a  $3+1$-dimensional anti-de Sitter setting, in section \ref{sec:Alternative-Scenarios-with}
we highlight some examples of viable physical pictures which are obtained by relaxing one or another of these
requirements.
We provide further guidance to the phenomenology of anti-de Sitter momentum space in Section \ref{sec:Magueijo-Smolin-Composition}, where we consider 
a much-studied DSR scenario, first introduced by Magueijo and Smolin
\cite{2002PhRvL..88s0403M}, and 
show that it is an example of  4D setting
adopting anti-de Sitter momentum space (whereas previous studies had erroneously linked it to
de Sitter momentum space), and that it is 
 a scenario such that only $R1$,
$R2$, $R4$ are enforced ($R3$ does not hold
in the scenario of Ref.\cite{2002PhRvL..88s0403M}).

\section{$2+1$ Gravity and $\kappa$-Poincar\'e momentum spaces \label{sec:Relative-Locality-Analysis}}
In this section we offer a brief summary of  previous results which
are relevant for our analysis. We start by offering a sketchy description of
the relative-locality framework, the framework which provides us
the tools for analyzing the properties of
relativistic kinematics in terms of geometric structures.
Then we discuss the results on 3D quantum gravity which are most relevant
for our analysis, and similarly we discuss some properties of the de Sitter momentum
space which admits description in terms of the $\kappa$-Poincar\'e Hopf algebra.
As we go along we shall characterize the requirements
$R1$,
$R2$, $R3$ and $R4$ in greater detail then already done in our
introductory remarks.

\subsection{The Relative Locality framework\label{sub:The-Relative-Locality}}
We fix notation by describing the on-shell relation and the
law of composition of momenta, in general, in the following form:
\begin{equation}
\begin{cases}
m^{2}=d_{\kappa}^{2}(k),\\
q=p\oplus_{\kappa}p',
\end{cases}\label{eq:The Laws (def)-1}
\end{equation}
where $q$ is the momentum obtained composing momenta $p$ and $p'$,
while $m$ is the mass of a particle of momentum $k$.
The functions $d_{\kappa}^{2}$ and $\oplus_{\kappa}$
depend on momenta and (in general) a deformation parameter
$\kappa$, a momentum scale often identified with the Planck scale
(we work in units with the speed-of-light scale and Planck's constant set to 1).

Within the relative-locality framework, introduced in \cite{principle},
one gives a geometrical picture of relativistic kinematics,
by associating the on-shell relation to a metric on momentum space
and the
composition law to a connection on momentum space, thereby characterizing momentum
space as a Cartan manifold:
\begin{equation}
\begin{array}{ccc}
d_{\kappa}^{2} & \leftrightarrow & g^{\mu\nu},\\
\oplus_{\kappa} & \leftrightarrow & \Gamma_{\lambda}^{\mu\nu}.
\end{array}
\end{equation}
\par
The on-shell relation is described
as the geodesic distance of a momentum $p$ from the origin of momentum space,
where the geodesic is evaluated according to the metric $g^{\mu\nu}$
on momentum space. This means that:
\begin{equation}
m^{2}=d_{\kappa}^{2}(p,0)=\int dt\sqrt{g^{\mu\nu}\dot{\gamma}_{\mu}\dot{\gamma}_{\nu}}\, .\label{eq:dispinterpretation}
\end{equation}
\par
For the association between  composition law and connection on momentum space
the original proposal was \cite{principle,balenascience}:
\begin{equation}
\Gamma_{\lambda}^{\mu\nu}(p)=-\frac{\partial}{\partial q_{\mu}}\frac{\partial}{\partial k_{\nu}}(q\oplus_{\kappa}^{[p]}k)_{\lambda}|_{q=k=p},\label{eq:standardcompinterpr}
\end{equation}
where $\oplus^{[p]}$ is such that
$q\oplus_{\kappa}^{[p]}k=p\oplus_\kappa\left(\ominus_{\kappa}p\oplus_{\kappa}q\right)\oplus_{\kappa}\left(\ominus_{\kappa}p\oplus_{\kappa}k\right)$
and $\ominus_{\kappa}$ is the antipode operation, such that $p\oplus_{\kappa}\left(\ominus_{\kappa}p\right)=0$.

Some of us advocated \cite{balenascience} an alternative association
between composition law and connection on momentum space
such that to any pair of momenta $q$ and $k$
one associates the autoparallel\footnote{An autoparallel curve $\gamma$ is such
that $\frac{d^{2}\gamma_{\lambda}}{dt}
+\Gamma_{\lambda}^{\mu\nu}\left(\gamma\right)\frac{d\gamma_{\mu}}{dt}\frac{d\gamma_{\nu}}{dt}=0$.} curves $\gamma^{(q)}$ and $\gamma^{(k)}$ which connect $q$ and $k$ to the
origin of momentum space. A parametric surface $\gamma\left(s,t\right):[0,1]\times[0,1]\rightarrow M$
function of the parameters $s$ and $t$ is then defined on the momentum
space $M$ by:
\begin{equation}
\begin{cases}
\frac{d}{dt}\frac{d}{ds}\gamma_{\lambda}+\Gamma_{\lambda}^{\mu\nu}\left(\gamma\right)\frac{d\gamma_{\mu}}{dt}\frac{d\gamma_{\nu}}{ds}=0,\\
\gamma_{\lambda}\left(s,0\right)=\gamma_{\lambda}^{(k)}\left(s\right),\\
\gamma_{\lambda}\left(0,t\right)=\gamma_{\lambda}^{(q)}\left(t\right).
\end{cases}\label{eq:cycliccompinterpr}
\end{equation}
The momentum composed of $q$ and $k$ is the point of momentum space
given by:
\begin{equation}
q\oplus_{\kappa}k=\gamma\left(1,1\right).
\end{equation}

In most cases of interest, and  in particular when the composition law is associative,
the two ways to link a composition law with a connection
on momentum space are equivalent.

\subsection{$2+1$ Gravity momentum space\label{sub:-Gravity-Momentum}}

As we mentioned in the introduction,
in $2+1$ dimensions the description of the interactions between
particles and the gravitational field is such that
\cite{freidellivineeffective,MatschullWelling}
 the
gravitational degrees of freedom can be integrated out, thereby providing an effective
description of the particles in which particle momenta live
on a curved space, which turns out to be the Lie group SL$\left(2,\mathbb{R}\right)$.

Any element of SL$\left(2,\mathbb{R}\right)$ can be written as \cite{2014SIGMA..10..079A}:
\begin{equation}
\hat{p}=\frac{p_{4}}{\kappa}\hat{I}-2\frac{p_{\mu}}{\kappa}\hat{X}^{\mu},\label{eq:momentum representation}
\end{equation}
where the hat denotes matrix elements, $p_4$ and $p_{\mu}$ $(\mu=0,1,2)$ are coefficients,
$\hat{I}$ is the identity matrix and the $\hat{X}^{\mu}$ belongs
to the Lie algebra of SL$\left(2,\mathbb{R}\right)$, for example
they can be chosen as:
\begin{equation}
\begin{array}{ccc}
\hat{X}^{0}=\frac{1}{2}\left(\begin{array}{cc}
0 & 1\\
-1 & 0
\end{array}\right), & \hat{X}^{1}=\frac{1}{2}\left(\begin{array}{cc}
0 & -1\\
-1 & 0
\end{array}\right), & \hat{X}^{2}=\frac{1}{2}\left(\begin{array}{cc}
1 & 0\\
0 & -1
\end{array}\right).\end{array}
\end{equation}
\par
The fact that $\hat{p}$ belongs to SL$\left(2,\mathbb{R}\right)$
means that its determinant has to be $1$ and this translates into
the condition:
\begin{equation}
p_{4}^{2}+p_{\mu}p^{\mu}=\kappa^2.
\end{equation}

The coefficients $p_{4}$ and $p_{\mu}$ can then be interpreted as
the coordinates of a curved momentum space with an anti-de Sitter
structure.
\par
The on-shell relation can be constructed by taking
the Killing form of the Lie algebra of SL$\left(2,\mathbb{R}\right)$ \cite{RellocinAdS}
and takes the form:
\begin{equation}
d_{AdS}\left(p,0\right)=\begin{cases}
\kappa \arccos\left(p_{4}/\kappa\right)\ ; & p_{4}<\kappa\, ,\\
0\ ; & p_{4}=\kappa\, ,\\
\kappa \arccosh\left(p_{4}/\kappa\right)\ ; & p_{4}>\kappa\, ,
\end{cases}\label{eq:AdSdistance-1-1-1}
\end{equation}
where $\kappa$ is usually identified with the Planck mass.
The composition law $p\oplus_{\kappa}q$ can be
introduced in terms of \cite{MatschullWelling,RellocinAdS}
\begin{equation}
\hat{p}\cdot\hat{q}=\kappa^{-1}\left(p\oplus_{\kappa}q\right)_{4}\hat{I}-2\kappa^{-1}\left(p\oplus_{\kappa}q\right)_{\mu}\hat{X}^{\mu},\label{eq:comp definition}
\end{equation}
where the dot stands for matrix multiplication.

From (\ref{eq:momentum representation}) and (\ref{eq:comp definition})
we get:
\begin{equation}
\begin{cases}
\left(q\oplus_{\kappa}p\right)_{\lambda}=\frac{1}{\kappa}\left(q_{\lambda}p_{4}+q_{4}p_{\lambda}-\epsilon_{\lambda}^{\ \alpha\beta}q_{\alpha}p_{\beta}\right),\\
\left(q\oplus_{\kappa}p\right)_{4}=\frac{1}{\kappa}\left(q_{4}p_{4}-q_{\alpha}p^{\alpha}\right).
\end{cases}\label{eq:2+1Gcomp}
\end{equation}

We adopt the Cartesian parameterization of anti-de Sitter, such that
\begin{equation}
p_{4}=\begin{cases}
\sqrt{\kappa^{2}-p^{2}}\ ;\  & p_{4}>0\, ,\\
-\sqrt{\kappa^{2}-p^{2}}\ ;\  & p_{4}<0\, ,
\end{cases}
\end{equation}
where $p^{2}=p_{\mu}p^{\mu}$ . Then the induced metric reads
\begin{equation}
g^{\mu\nu}\left(p\right)=\eta^{\mu\nu}-\frac{p^{\mu}p^{\nu}}{p^{2}-\kappa^{2}}\, ,
\end{equation}
and the associated geodesic distance is given by:
\begin{equation}
d_{AdS}\left(q,k\right)=\begin{cases}
\kappa \arccos\left[\left(q_{4}k_{4}+q_{\mu}k^{\mu}\right)/\kappa^{2}\right]\ ; & q_{4}k_{4}+q_{\mu}k^{\mu}<\kappa^2\, ,\\
0\ ; & q_{4}k_{4}+q_{\mu}k^{\mu}=\kappa^2\, ,\\
\kappa \arccosh\left[\left(q_{4}k_{4}+q_{\mu}k^{\mu}\right)/\kappa^{2}\right]\ ; & q_{4}k_{4}+q_{\mu}k^{\mu}>\kappa^2\, .
\end{cases}\label{eq:AdSdistance-1-1}
\end{equation}
By picking as origin the point $p_{4}=\kappa$, $p_{\mu}=0$, the on-shell
relation can be obtained from (\ref{eq:dispinterpretation}) and reduces
to (\ref{eq:AdSdistance-1-1-1}).

For what concerns the connection, since the composition law is associative
one can rely equivalently on (\ref{eq:standardcompinterpr})
or (\ref{eq:cycliccompinterpr}), obtaining
\begin{equation}
\Gamma_{\lambda}^{\mu\nu}\left(0\right)=\frac{1}{\kappa}\epsilon_{\lambda}^{\ \mu\nu}\, ,
\end{equation}
It is easy to see that the on-shell relation and the composition law
are both invariant under the (undeformed) group of Lorentz transformations, $p \rightarrow \Lambda(p)$:
\begin{equation}
\begin{cases}
d_{\kappa}\left(\Lambda\left(p\right),0\right)=d_{\kappa}\left(p,0\right),\\
\Lambda\left(q\right)\oplus_{\kappa}\Lambda\left(k\right)=\Lambda\left(q\oplus_{\kappa}k\right).
\end{cases}\label{eq:2+1Gcovariance}
\end{equation}


\subsection{$\kappa$-Poincar\'e momentum space\label{sub:-Poincar=0000E9-Momentum-Space}}

The $\kappa$-Poincar\'e momentum space is the most studied scenario
for deformed relativistic kinematics.
It can be characterized by the following on-shell relation and law of composition
of momenta \cite{KowalskiGlikman:2004tz}:
\begin{equation}
\begin{cases}
d_{\kappa}\left(\tilde{p}\right)=\kappa \arccosh\left[\cosh\left(\tilde{p}_{0}/\kappa\right)-\frac{1}{2\kappa^{2}}\tilde{p}_{1}^{2}e^{p_{0}/\kappa}\right],\\
\left(\tilde{q}\oplus_{\kappa}\tilde{k}\right)_{0}=\tilde{q}_{0}+\tilde{k}_{0}\, ,\\
\left(\tilde{q}\oplus_{\kappa}\tilde{k}\right)_{1}=\tilde{q}_{1}+\tilde{k}_{1}e^{-\tilde{q}_{0}/\kappa}\, ,
\end{cases}
\end{equation}
where for simplicity we are giving explicit formulas for the case of $1+1$ spacetime dimensions.
The
on-shell relation can be associated \cite{GiuliaFlavio} to a de Sitter
metric in the so called comoving coordinates (see appendix \ref{appendix_anti1}):
\begin{equation}
g^{\mu\nu}=\left(\begin{array}{cc}
1 & 0\\
0 & -e^{-\tilde{p}_{0}/\kappa}
\end{array}\right),
\end{equation}
and the composition law can be associated to a connection of the form:
\begin{equation}
\Gamma_{\lambda}^{\mu\nu}\left(\tilde{p}\right)=\frac{1}{\kappa}\delta_{\lambda}^{1}\delta_{0}^{\mu}\delta_{1}^{\nu}\, .
\end{equation}
This connection is torsionful and its  curvature  vanishes.
\par
The $\kappa$-Poincar\'e on-shell relation is invariant under the following deformation
of Lorentz transformations:
\begin{align}
\tilde{\Lambda}_\kappa\left(\xi,\tilde{p}\right)=\begin{cases}
\tilde{\Lambda}{}_{0}\left(\tilde{p}\right)=&\kappa\ln\left\{\frac{1}{\kappa}\left[\cosh\left(\xi\right)\left[\kappa\sinh\left(\tilde{p}_{0}/\kappa\right)+\frac{1}{2\kappa}\tilde{p}_{1}^{2}e^{\tilde{p}_{0}/\kappa}\right]+\sinh\left(\xi\right)e^{\tilde{p}_{0}/\kappa}\tilde{p}_{1}\right.\right.\\
&\left.\left.+\kappa\cosh\left(\tilde{p}_{0}/\kappa\right)-\frac{1}{2\kappa}\tilde{p}_{1}^{2}e^{\tilde{p}_{0}/\kappa}\right]\right\}\, ,\\
\tilde{\Lambda}{}_{1}\left(\tilde{p}\right)=&\kappa\frac{\cosh\left(\xi\right)e^{\tilde{p}_{0}/\kappa}\tilde{p}_{1}+\sinh\left(\xi\right)\left[\kappa\sinh\left(\tilde{p}_{0}/\kappa\right)+\frac{1}{2\kappa}\tilde{p}_{1}^{2}e^{\tilde{p}_0/\kappa}\right]}{\sinh\left(\xi\right)e^{\tilde{p}_{0}/\kappa}\tilde{p}_{1}+\cosh\left(\xi\right)\left[\kappa\sinh\left(\tilde{p}_{0}/\kappa\right)+\frac{1}{2\kappa}\tilde{p}_{1}^{2}e^{\tilde{p}_{0}/\kappa}\right]+\kappa\cosh\left(\tilde{p}_{0}/\kappa\right)-\frac{1}{2\kappa}\tilde{p}_{1}^{2}e^{\tilde{p}_{0}/\kappa}}\, ,
\end{cases}\label{eq:kappaboost}
\end{align}
where $\xi$ is the rapidity parameter.

The composition law is covariant in a non-standard way:  the action of the deformed
boosts (\ref{eq:kappaboost}) on the composition law is such that:
\begin{equation}
\tilde{\Lambda}_{\kappa}\left(q\oplus_{\kappa}k,\xi\right)=\tilde{\Lambda}_{\kappa}\left(q,\xi\right)\oplus_{\kappa}\tilde{\Lambda}_{\kappa}\left(k,\xi'\left(q\right)\right),
\end{equation}
where $\xi'\left(q\right)$ is a function of $q$ of the form
\begin{equation}
\xi'=2\arcsinh\left(\frac{e^{-\tilde{p}_{0}/\kappa}\sinh\left(\xi/2\right)}{\sqrt{\left(\cosh\left(\xi/2\right)+\frac{1}{\kappa}\tilde{p}_{1}\sinh\left(\xi/2\right)\right)^{2}-e^{-2\tilde{p}_{0}/\kappa}\sinh^{2}\left(\xi/2\right)}}\right).
\end{equation}
We shall refer in the following to such kind of behavior as a ``non-homomorphic
action'' of the Lorentz boost on the composition law.
Notice that the covariance of the $\kappa$-Poincar\'e composition law is non-standard in
two ways, some of which are encoded in  the form of
(\ref{eq:kappaboost}), while others are encoded in the fact that boosts
act with a non-homomorphic action on the composition law.
Still, as shown in Refs.\cite{balenascience,tesipalmisano}, it is a fully legitimate
(logically consistent) way of enforcing relativistic covariance.

\subsection{Some key properties of $2+1$ gravity and $\kappa$-Poincar\'e momentum space\label{sub:Properties-of-}}
It is useful for our purposes to highlight a few key properties that are found
both for the composition law of  $2+1$ gravity and for the
composition law of the $\kappa$-Poincar\'e momentum space. We start from
property $R1$, the composition laws admit an identity element, such that
\begin{equation}
\begin{array}{cc}
R1: & \begin{cases}
q\oplus_{\kappa}p|_{p=0}=q\, ,\\
q\oplus_{\kappa}p|_{q=0}=p\, .
\end{cases}\end{array}\label{eq:IdentityExistence}
\end{equation}

The composition laws of  $2+1$ gravity and of the $\kappa$-Poincar\'e momentum space
 also have properties $R2$ and $R3$, {\it i.e.} they preserve the geodesic distance in momentum space,
\begin{equation}
\begin{array}{cc}
R2: & d_{\kappa}\left(p\oplus_{\kappa}q,p\oplus_{\kappa}k\right)=d_{\kappa}\left(q,k\right)\, ;\end{array}\label{eq:Isometricity}
\end{equation}
and they are associative,
\begin{equation}
\begin{array}{cc}
R3: & \left(p\oplus_{\kappa}\left(q\oplus_{\kappa}k\right)\right)_{\lambda}=\left(\left(p\oplus_{\kappa}q\right)\oplus_{\kappa}k\right)_{\lambda}\, .\end{array}\label{eq:Associativity}
\end{equation}

For what concerns property $R4$ (relativistic covariance)
we have seen
in subsections \ref{sub:-Gravity-Momentum} and \ref{sub:-Poincar=0000E9-Momentum-Space}
that the compatibility with a relativistic description is encoded in the
following covariance properties:
\begin{equation}
\begin{array}{cc}
 & \begin{cases}
2+1\ \text{Gravity} & \Lambda\left(q\oplus_{\kappa}k\right)=\Lambda\left(q\right)\oplus_{\kappa}\Lambda\left(k\right)\, ,\\
\mbox{\ensuremath{\kappa}}\text{-Poincar\'e} & \tilde{\Lambda}_{\kappa}\left(q\oplus_{\kappa}k,\xi\right)
=\tilde{\Lambda}_{\kappa}\left(q,\xi\right)\oplus_{\kappa}
\tilde{\Lambda}_{\kappa}\left(k,\xi'\left(q\right)\right)\, .
\end{cases}\end{array}
\end{equation}
\par
In general, in a relativistic momentum space
there will be a deformation $\tilde{\Lambda}\left(p,\xi\right)$
of the Lorentz transformations,  and suitable functions $\xi'\left(\xi,p\right)$
and $\xi''\left(\xi,p\right)$ such that under the action of $\tilde{\Lambda}$
the on-shell relation is invariant
\begin{equation}
d_{\kappa}\left(\tilde{\Lambda}\left(p\right)\right)=d_{\kappa}\left(p\right)\, ,
\end{equation}
and the composition law is covariant, meaning that
\begin{equation}
\begin{array}{cc}
R4: & \tilde{\Lambda}_{\kappa}\left(q\oplus_{\kappa}k,\xi\right)=\tilde{\Lambda}_{\kappa}\left(q,\xi'\left(\xi,k\right)\right)\oplus_{\kappa}\tilde{\Lambda}_{\kappa}\left(k,\xi''\left(\xi,q\right)\right)\, .\end{array}\label{eq:covariance}
\end{equation}
The triple $\left(\tilde{\Lambda},\xi',\xi''\right)$
should be global, meaning that for any momentum $p$ and any
given finite $\xi$ we have that both $\tilde{\Lambda}\left(\xi,q\right)$,
and $\tilde{\Lambda}\left(q,\xi'\left(\xi,k\right)\right)\oplus\tilde{\Lambda}\left(k,\xi''\left(\xi,q\right)\right)$
are finite.

In the following, a composition law satisfying the properties (\ref{eq:IdentityExistence})-(\ref{eq:covariance})
with respect to some metric will be called a ``Hopf composition law''
associated to the given metric.
We label as ``Hopf'' a composition law satisfying the properties (\ref{eq:IdentityExistence})-(\ref{eq:covariance}),
since we shall conjecture a link between momentum spaces governed by
(\ref{eq:IdentityExistence})-(\ref{eq:covariance}),
and the mathematics of Hopf
algebras.

One last point we want to remark here concerns the requirement of
covariance and particularly the possible distinction between the general
realization of covariance expressed in (\ref{eq:covariance}) and
the requirement of homomorphic covariance, {\it i.e.}
\begin{equation}\label{R4b}
\begin{array}{cc}
R4b: & \tilde{\Lambda}\left(q\oplus_{\kappa}k,\xi\right)=\tilde{\Lambda}\left(q,\xi\right)\oplus_{\kappa}\tilde{\Lambda}\left(k,\xi\right)\, ,\end{array}
\end{equation}
where we can see that the parameters identifying the Lorentz transformations
are all identical on both sides of the equation.

Although the distinction between the homomorphic and the non-homomorphic
realizations of covariance are useful in classifying the possible
structures of momentum spaces, we stress that both cases constitute
a legitimate implementation of the relativity principle, as discussed
in \cite{balenascience,tesipalmisano}.

We notice that for maximally symmetric momentum spaces parametrized
in Cartesian coordinates, the on-shell relation is preserved by the
standard undeformed Lorentz transformations $\Lambda$ and the requirement
of homomorphic covariance is equivalent to the condition that the composition
law can be expressed in the explicitly covariant form:
\begin{equation}
\left(q\oplus_{\kappa}k\right)_{\lambda}=A\left(q,k\right)q_{\lambda}+B\left(q,k\right)k_{\lambda}\, ,
\end{equation}
where $A$ and $B$ are scalar functions of $q$ and $k$ .
\par
On the other hand, the requirement of covariance in the more general
form (\ref{eq:covariance}), is equivalent to the condition that:
\begin{equation}
\left(q\oplus_{\kappa}k\right)_{\lambda}=A\left(q,k\right)\chi_{\lambda}^{\ \alpha}q_{\alpha}+B\left(q,k\right)\Omega_{\lambda}^{\ \alpha}k_{\alpha}\, ,
\end{equation}
where $\chi$ and $\Omega$ are now arbitrary Lorentz transformations,
functions of $q,$ $k$ and any other fixed arbitrary momenta. For
example, introducing a single external vector $v$ we can consider
Lorentz transformations of the form:
\begin{equation}
\Omega_{\lambda}^{\ \alpha}=\omega_{1}\left(q,v\right)\delta_{\lambda}^{\alpha}+\omega_{2}\left(q,v\right)q_{\lambda}q^{\alpha}+\omega_{3}\left(q,v\right)q_{\lambda}v^{\alpha}+\omega_{4}\left(q,v\right)q_{\lambda}q^{\alpha}+\omega_{5}\left(q,v\right)\epsilon_{\lambda}^{\ \alpha\beta\gamma}q_{\beta}v_{\gamma}\, ,
\end{equation}
where $\omega_{n}$ ($n=1,..., 5$) are scalar functions of $q$
and $v$.\footnote{Of course the scalars $\omega_{n}$ are not independent;
they have to be compatible with the (pseudo) orthogonality conditions
for $\Omega$.}
\par
The compatibility between the introduction of an external vector and
the covariance of the composition law will be discussed in subsection
\ref{sub:Aside-on-the}.

\subsection{``Proper de Sitter'' and ``Snyder-de Sitter'' momentum spaces\label{sub:Proper-and-Snyder}}

While most of the work
in the context of deformed relativistic kinematic has been developed
in ways that establish a consistency with the formalism of  Hopf algebras,
 it is important for our purposes to stress that
 alternatives to this ``Hopf-algebra paradigm''
 can also be legitimately considered,
 and in fact they have been already considered in the literature,
 most notably in Refs.\cite{balenascience,2014PhRvD..90g6010B}.
 These are two cases of momentum spaces with
 de Sitter metric not admitting a Hopf-algebraic description,
 and therefore can serve as reference models for the investigations
 we shall also report here on
 momentum spaces with
 anti-de Sitter metric not admitting a Hopf-algebraic description.

 We describe both the model of Ref.\cite{balenascience}
 and the model of Ref.\cite{2014PhRvD..90g6010B}
 adopting Cartesian coordinates on de Sitter momentum space.\footnote{In \cite{balenascience} Proper de Sitter momentum space has been
introduced and developed in comoving coordinates (see the appendix
\ref{appendix_anti1}).} These can be derived from the embedding coordinates (see the appendix
\ref{appendix_anti1}) by imposing:
\begin{equation}
p_{4}=\begin{cases}
\sqrt{\kappa^{2}+p^{2}}\ ;\  & p_{4}>0\, ,\\
-\sqrt{\kappa^{2}+p^{2}}\ ;\  & p_{4}<0\, .
\end{cases}\label{eq:Cartesian coord 2}
\end{equation}

In particular, both in Ref.\cite{balenascience} and in Ref.\cite{2014PhRvD..90g6010B}
work within the $p_{4}>0$ sector of de Sitter momentum space.
The induced metric is
\begin{equation}
g^{\mu\nu}\left(p\right)=\eta^{\mu\nu}-\frac{p^{\mu}p^{\nu}}{p^{2}+\kappa^{2}}\, ,
\end{equation}
and the associated on-shell relation is
\begin{equation}
d_{\kappa}^{2}\left(p\right)=\kappa^{2}\arccosh^{2}\left(\sqrt{p^{2}+\kappa^{2}}/\kappa\right)\, .\label{eq:proper snyder on-shell relations}
\end{equation}

The ``Proper de Sitter'' composition law of Ref.\cite{balenascience}
is defined as the composition
law (\ref{eq:cycliccompinterpr}) associated to the Levi Civita de
Sitter connection that in Cartesian coordinates is given by $\Gamma_{\lambda}^{\mu\nu}\left(p\right)=\frac{1}{\kappa}p_{\lambda}g^{\mu\nu}\left(p\right)$.

At second order in the Planck length $\kappa^{-1}$ it is possible \cite{balenascience}
to give a general solution to the equations (\ref{eq:cycliccompinterpr})
in the form
\begin{align}
\left(q\oplus_{\kappa}p\right)_{\lambda}=q_{\lambda}+k_{\lambda}-\frac{1}{\kappa}\Gamma_{\lambda}^{\alpha\beta}\left(0\right)q_{\alpha}k_{\beta}-\frac{1}{\kappa^{2}}\left\{ \left(\partial^{\rho}\Gamma_{\lambda}^{\alpha\beta}\left(0\right)\right)q_{\alpha}k_{\beta}\left(\frac{q_{\rho}+k_{\rho}}{2}\right)\right.\nonumber\\
\left.-\frac{1}{2}\left(\Gamma_{\lambda}^{\alpha\beta}\Gamma_{\alpha}^{\gamma\delta}\right)\left(0\right)q_{\gamma}k_{\delta}k_{\beta}-\frac{1}{2}\left(\Gamma_{\lambda}^{\alpha\beta}\Gamma_{\beta}^{\gamma\delta}\right)\left(0\right)q_{\alpha}q_{\gamma}k_{\delta}\right\}\, . \label{eq:polyakovsecondorder}
\end{align}
Substituting the Levi Civita connection into (\ref{eq:polyakovsecondorder})
one gets
\begin{equation}
\left(q\oplus_{\kappa}p\right){}_{\lambda}=\left(q_{\lambda}+k_{\lambda}\right)\left(1-\frac{1}{2\kappa^{2}}q\cdot k\right)\, .\label{eq:Proper comp}
\end{equation}

We notice that the composition law (\ref{eq:Proper comp})
satisfies $R1$ and is commutative but not associative. Furthermore
it satisfies $R4b$ with respect to the standard Lorentz boosts:
\begin{equation}
\Lambda\left(q\oplus_{\kappa}k\right)=\Lambda\left(q\right)\oplus_{\kappa}\Lambda\left(k\right)\, .\label{eq:proper covariance}
\end{equation}
\par
The ``Snyder'' composition law of Ref.\cite{2014PhRvD..90g6010B}
is 
\begin{equation}
\begin{cases}
(p\oplus q)_{\mu}=\frac{1}{\kappa}p_{\mu}q_{4}+\left[\delta_{\mu}^{\alpha}+\frac{p_{\mu}p^{\alpha}}{\kappa(\kappa+p_{4})}\right]q_{\alpha}\, ,\\
(p\oplus q)_{4}=\frac{1}{\kappa}\left(p_{4}q_{4}+p_{\mu}q^{\mu}\right)\
\end{cases}
\end{equation}
It satisfies the requirements $R1$, $R2$ and $R4b$ (again, with respect to the standard Lorentz
boost) and is non-associative and non-commutative.

\section{$3+1$-dimensional anti-de Sitter momentum spaces\label{sec:-Dimensional-anti-de}}
In the previous section we have highlighted four properties
for the laws of composition of momenta
in $2+1$ gravity and
in the $\kappa$-Poincar\'e momentum space:
$R1$ (existence of identity), $R2$ (isometricity),
$R3$ (associativity),
and $R4$ (relativistic covariance, with variant $R4b$ for homomorphic covariance).

In this section we will show that these four properties $R1$-$R4$
enjoyed by the laws of composition of momenta
in $2+1$ gravity and
in the $\kappa$-Poincar\'e momentum space,
cannot be enforced in the case of the $3+1$-dimensional anti-de Sitter
momentum space.

\subsection{Geodesic distance and on-shell relation}

We start by identifying the form of the geodesic distance in $3+1$
dimensional anti-de Sitter momentum space. The anti-de Sitter manifold
can be characterized as a hypersurface, embedded in a $3+2$ dimensional
flat space, characterized by:
\begin{equation}
p_{0}^{2}-p_{1}^{2}-p_{2}^{2}-p_{3}^{2}+p_{4}^{2}=\kappa^{2}\, .
\end{equation}

The geodesic distance between any pair of momenta $q$ and $k$
is:
\begin{equation}
d_{AdS}\left(q,k\right)=\begin{cases}
\kappa\arccos\left[\left(q_{4}k_{4}+q_{\mu}k^{\mu}\right)/\kappa^{2}\right]\ ;\  & q_{4}k_{4}+q_{\mu}k^{\mu}<\kappa^2\, ,\\
0\ ;\  & q_{4}k_{4}+q_{\mu}k^{\mu}=\kappa^2\, ,\\
\kappa\arccosh\left[\left(q_{4}k_{4}+q_{\mu}k^{\mu}\right)/\kappa^{2}\right]\ ;\  & q_{4}k_{4}+q_{\mu}k^{\mu}>\kappa^2\, .
\end{cases}\label{eq:AdSdistance-1}
\end{equation}

As specified by the relative-locality framework \cite{principle},
the on-shell relation can be inferred from (\ref{eq:AdSdistance-1})
by choosing as origin of our momentum space the point $p_O=\left(0,0,0,0,\kappa\right)$,
obtaining \cite{Arzano:2014jua}:
\begin{equation}
m_{AdS}\left(p\right)=d_{AdS}\left(p,p_O\right)=\begin{cases}
\kappa \arccos\left(p_4/\kappa\right)\ ;\  & p_4<\kappa\, ,\\
0\ ;\  & p_4=\kappa\, ,\\
\kappa \arccosh\left(p_4/\kappa\right)\ ;\  & p_4>\kappa\, .
\end{cases}\label{eq:kappaAdSdispersion-1}
\end{equation}


\subsection{General isometric composition laws}
Our next task is to identify what are the isometric composition
laws, {\it i.e.} the composition laws which satisfy the first three properties
$R1$, $R2$, $R3$.

The requirement of invariance of the geodesic distance under the action
of the composition of momenta, characterizes the composition law
 as an element of $\text{SO}\left(2,3\right)/SO\left(1,3\right)$
which is the isometry group of anti-de Sitter quotiented by the subgroup
of isometries that preserve the origin we picked for the momentum
space. In particular it can be written as a linear action:
\begin{equation}
\left(p\oplus_{H}q\right)=T\left(p\right)\cdot q\, ,
\end{equation}
where $\oplus_H$ stands for ``Hopf composition law'', $q$ is a five dimensional vector and $T$ is a $5\times5$
matrix of $\text{SO}\left(2,3\right)$ and a suitable function of
$p$.\footnote{A similar approach has been recently explored in \cite{Carmona:2019vsh,Carmona:2019fwf,Relancio:2022mia,Carmona:2021pxw,Relancio:2021ahm,Carmona:2021gbg}.}

We can write $T$ in all generality as:
\begin{equation}
T\left(p\right)=\bar{T}\left(p\right)\Omega\left(p\right),\label{eq:ambiguity}
\end{equation}
where $\Omega\left(p\right)$
is an arbitrary Lorentz transformation and $\bar{T}$ is such that $R1$ and $R2$ are satisfied for any choice of $\Omega\left(p\right)$.

It is easy to see that the anti-de Sitter geodesic distance (\ref{eq:AdSdistance-1}) is preserved by the action of a composition
law of the form:
\begin{equation}
\left(p\oplus_{H}q\right)_{I}=\left[\frac{1}{\kappa}p_{I}\delta_{4}^{J}-\frac{1}{\kappa}p^{\alpha}\delta_{I}^{4}\delta_{\alpha}^{J}+\delta_{I}^{\lambda}\left(\delta_{\lambda}^{\alpha}-\frac{p_{\lambda}p^{\alpha}}{\kappa(\kappa+p_{4})}\right)\delta_{\alpha}^{J}\right]q_{J}\, ,\label{eq:barekappaAdScomp}
\end{equation}
or, in a more explicit form:
\begin{equation}
\begin{cases}
\left(p\oplus_{H}q\right)_{\lambda}=\frac{1}{\kappa}p_{\lambda}q_{4}+\left(\delta_{\lambda}^{\alpha}-\frac{p_{\lambda}p^{\alpha}}{\kappa(\kappa+p_{4})}\right)q_{\alpha}\, ,\\
\left(p\oplus_{H}q\right)_{4}=\frac{1}{\kappa}\left(p_{4}q_{4}-p_{\mu}q^{\mu}\right)\, .
\end{cases}\label{eq:kappaAdScomplawCartisian}
\end{equation}

A more general possibility, provided by the possible choices of $\Omega$ in (\ref{eq:ambiguity}),
has the form:
\begin{equation}
\begin{cases}
\left(p\oplus_{H}q\right)_{\lambda}=\frac{1}{\kappa}p_{\lambda}q_{4}+\left(\delta_{\lambda}^{\alpha}-\frac{p_{\lambda}p^{\alpha}}{\kappa(\kappa+p_{4})}\right)\Omega\left(p\right)_{\alpha}^{\ \beta}q_{\beta}\, ,\\
\left(p\oplus_{H}q\right)_{4}=\frac{1}{\kappa}\left(p_{4}q_{4}-p^{\alpha}\Omega\left(p\right)_{\alpha}^{\ \beta}q_{\beta}\right)\, .
\end{cases}\label{eq:generalkappaAdScomp-1}
\end{equation}
This is the general form of an anti-de Sitter isometric composition
law.

\subsection{Explicitly covariant isometric composition laws\label{sub:Explicitelly-Covariant-Isometric}}
Our next step is to look for a form of Lorentz transformations $\Omega$
such that the composition law is associative.
We first look for Lorentz transformations $\Omega$ with an explicitly covariant form. In $2+1$ dimensions such kind of transformations can be written in all generality as follows:
\begin{equation}
\Omega_{\alpha}^{\ \beta}=\omega_{1}\left(p\right)\delta_{\alpha}^{\beta}+\frac{1}{\kappa^{2}}\omega_{2}\left(p\right)p_{\alpha}p^{\beta}+\frac{1}{\kappa}\omega_{3}\left(p\right)\epsilon_{\alpha}^{\ \beta\gamma}p_{\gamma}\, ,
\end{equation}
where $\omega_{1}$, $\omega_{2}$ and $\omega_{3}$ are scalars that
are constrained by the orthonormality conditions:
\begin{equation}
\begin{cases}
\left(\omega_{3}\right)^{2}=\frac{\kappa^{2}}{p^{2}}\left[1-\left(\omega_{1}\right)^{2}\right]\, ,\\
2\left(\omega_{1}\right)\left(\omega_{2}\right)+\left(\omega_{2}\right)^{2}\frac{p^{2}}{\kappa^{2}}-\frac{\kappa^{2}}{p^{2}}\left[1-\left(\omega_{1}\right)^{2}\right]=0\, .
\end{cases}
\end{equation}
Evidently one of the scalars $\omega_{1}$, $\omega_{2}$
or $\omega_{3}$ can be arbitrarily chosen, and in particular the choice
\begin{equation}
\begin{array}{ccc}
\omega_{1}=\frac{p_{4},}{\kappa}, & \omega_{2}=\frac{\kappa}{\kappa+p_{4}}, & \omega_{3}=-1\end{array}
\end{equation}
leads to the associative composition law (\ref{eq:2+1Gcomp}) arising
in $2+1$ gravity, as one can easily check.

In the four dimensional case, on the other hand, the most general
explicitly covariant Lorentz transformation has the form:
\begin{equation}
\Omega_{\alpha}^{\ \beta}=\omega_{1}\left(p\right)\delta_{\alpha}^{\beta}+\frac{1}{\kappa^{2}}\omega_{2}\left(p\right)p_{\alpha}p^{\beta}\, ,\label{eq:covariant3+1omega}
\end{equation}
where the scalar functions $\omega_{1}$ and $\omega_{2}$ are fixed
by the orthonormality conditions to be $\omega_{1}=1$ and $\omega_{2}$
either zero or equal to $-\frac{2\kappa^{2}}{p^{2}}$. The arising
composition laws would have then the two possible forms:
\begin{equation}
\begin{cases}
1) & \begin{cases}
\left(p\oplus q\right)_{\lambda}=\frac{1}{\kappa}p_{\lambda}q_{4}+\left(\delta_{\lambda}^{\alpha}-\frac{p_{\lambda}p^{\alpha}}{\kappa(\kappa+p_{4})}\right)q_{\alpha}\, ,\\
\left(p\oplus q\right)_{4}=\frac{1}{\kappa}\left(p_{4}q_{4}-p^{\alpha}q_{\alpha}\right)\, .
\end{cases}\\
2) & \begin{cases}
\left(p\oplus q\right)_{\lambda}=\frac{1}{\kappa}p_{\lambda}q_{4}+\left(\delta_{\lambda}^{\alpha}-\frac{p_{\lambda}p^{\alpha}}{\kappa(\kappa-p_{4})}\right)q_{\alpha}\, ,\\
\left(p\oplus q\right)_{4}=\frac{1}{\kappa}\left(p_{4}q_{4}+p^{\alpha}q_{\alpha}\right)\, ,
\end{cases}
\end{cases}\label{eq:AntiSnyder comp laws}
\end{equation}
which remind us of the Snyder composition law discussed in subsection \ref{sub:Proper-and-Snyder}.

We notice that both the composition laws in (\ref{eq:AntiSnyder comp laws})
are not associative. Therefore we conclude that in $3+1$ dimensions
it is not possible to find a composition law satisfying $R1$, $R2$,
$R3$ and $R4b$. 
This is consistent with and generalizes the result reported in the appendix of 
\cite{Carmona:2019fwf},
which showed the incompatibility of 
the conditions $R1$-$R4$. Our observation is stronger since we also contemplated replacing $R4$ with the weaker $R4b$,
and still found incompatibility. From the phenomenology perspective it is important to notice that the main difference between $R4$ and $R4b$
is that $R4$ assumes spatial isotropy, not assumed in $R4b$.
In the appendix of 
\cite{Carmona:2019fwf},
it was 
it is shown that, assuming isotropy, it is not possible to find a Lie subalgebra of translation generators in four dimensional anti-de Sitter space, which in turn indeed implies that there is no isotropic, associative, covariant composition law defined from the translation generators in four-dimensional anti-de Sitter space, {\it i.e.} the conditions $R1$-$R4$.

The main goal of the study we are here reporting, to which we devote the remaining subsections of this section, is to provide guidance to the phenomenology of 4D anti-de Sitter momentum space, taking as starting point
the incompatibility between the requirements 
 $R1$, $R2$,
$R3$ and $R4$ (or $R4b$). 
If there had been compatibility between these requirements then that scenario would have been a natural first target for the phenomenology. That not being the case (and the option of exploring in full generality the phenomenology of anti-de Sitter momentum space not being an available option, considering the smallness of the Planck length) it is valuable to identify some target scenarios which might deserve priority in phenomenology, and of course renounce to one or another of the conditions 
 $R1$, $R2$,
$R3$ and $R4$(/$R4b$).


\subsection{Aside on the possibility to describe general implementations
of the relativity principle\label{sub:Aside-on-the}}
 The results of the previous subsection on the impossibility of finding
a composition law satisfying $R1$, $R2$,
$R3$ and $R4b$, mean that in $3+1$ dimensions, with anti-de Sitter momentum space,
 one cannot reproduce exactly the structure encountered in 2+1-dimensional quantum gravity.
 Our next task is to establish whether or not it is at least possible to trade $R4{b}$ for the requirement of relativistic  covariance $R4$. This is inspired by the fact that
 in the de Sitter case by enforcing $R4$ (but not $R4b$) one finds the interesting
  case of the $\kappa$-Poincar\'e momentum space.

Once we drop the request of explicit covariance, we can choose the
Lorentz transformation $\Omega$ with a more general form than the
one we considered in (\ref{eq:covariant3+1omega}). The possibility
that we will consider consists in the introduction of an external
vector $v$, that we call the stabilizer, in terms of which $\Omega$
can be written, in all generality as:
\begin{align}
\Omega_{\alpha}^{\ \beta}=\omega_{1}\left(p^{2},v\right)\delta_{\alpha}^{\beta}+\frac{1}{\kappa^{2}}\omega_{2}\left(p^{2},v\right)p_{\alpha}p^{\beta}+\frac{1}{\kappa}\omega_{3}\left(p^{2},v\right)\epsilon_{\alpha}^{\ \beta\gamma\delta}p_{\gamma}v_{\delta}\nonumber\\
+\frac{1}{\kappa}\omega_{4}\left(p^{2},v\right)v_{\alpha}p^{\beta}+\frac{1}{\kappa}\omega_{5}\left(p^{2},v\right)p_{\alpha}v^{\beta}\, ,\label{eq:generalLorentzW}
\end{align}
where $\epsilon_{\alpha}^{\ \beta\gamma\delta}$ is the Levi Civita symbol and the scalars $\omega_{n}$ are constrained
by the orthonormality conditions:
\begin{equation}
\begin{cases}
\left(\omega_{1}\right)^{2}+\frac{1}{\kappa^{2}}\left(\omega_{3}\right)^{2}\left(p^{2}v^{2}-\left(p\cdot v\right)^{2}\right)=0\, ,\\
\left(\omega_{5}\right)^{2}=\left(\omega_{3}\right)^{2}\, ,\\
\frac{1}{\kappa}\left(p\cdot v\right)\left(\omega_{3}\right)^{2}+\omega_{1}\omega_{4}+\omega_{1}\omega_{5}+\frac{1}{\kappa^{2}}\omega_{2}\omega_{5}p^{2}+\frac{1}{\kappa}\omega_{4}\omega_{5}\left(v\cdot p\right)=0\, ,\\
2\omega_{1}\omega_{2}+\frac{1}{\kappa^{2}}\left(\omega_{2}\right)^{2}p^{2}+\left(\omega_{4}\right)^{2}v^{2}+\frac{1}{\kappa}2\omega_{4}\omega_{2}\left(v\cdot p\right)-\left(\omega_{3}\right)^{2}v^{2}=0\, .
\end{cases}\label{eq:Ortho Constraints}
\end{equation}

In the next subsection we will discuss whether it is possible to choose the
scalars $\omega_{n}$ ($n=1,...,5$) compatibly with (\ref{eq:Ortho Constraints}),
and in such a way that the composition law is associative. First here
we discuss how the introduction of an
external vector $v$ can be compatible with the overall DSR relativistic
compatibility of the composition law. As it has been shown in \cite{tesipalmisano},
the key point is given by the fact that $v$ enters the composition
law just inside the Lorentz transformations $\Omega$. Standard properties
of the Lorentz transformations ensure that for any such transformation
$\Lambda_{\xi}$ associated to a rapidity parameter $\xi$, we have:
\begin{equation}
\Omega\left(\Lambda_{\xi}q;v\right)=\left(\Lambda_{\xi}\Omega\left(q;v\right)\bar{\Lambda}_{\xi}\right)\bar{\Lambda}\left(\xi'\right)\, ,\label{eq:gen cov of Om-1}
\end{equation}
where $\bar{\Lambda}$ is the inverse of $\Lambda$ and $\xi'$ is
a set of parameters, function of $q$ and $v$, defined by:\footnote{The possibility to find the explicit solution for $\xi'$ is ensured
by the fact that the term $\Lambda_{\xi}\bar{\Omega}\left(q;v\right)\Omega\left(q;\bar{\Lambda}_{\xi}v\right)\bar{\Lambda}_{\xi}$
is a combination of Lorentz transformations and it is then also a
Lorentz transformation.}
\begin{equation}
\bar{\Lambda}\left(\xi'\right)=\Lambda_{\xi}\bar{\Omega}\left(p;v\right)\Omega\left(p;\bar{\Lambda}_{\xi}v\right)\bar{\Lambda}_{\xi}\, .
\end{equation}
By using (\ref{eq:gen cov of Om-1}) it is then easy to verify that
the composition law (\ref{eq:generalkappaAdScomp-1}) satisfies the
generalized form of covariance:
\begin{equation}
\Lambda_{\xi}\left(q\oplus_{H}k\right)=\Lambda_{\xi}\left(q\right)\oplus_{H}\Lambda_{\xi'}\left(k\right)\, .\label{eq:back cov}
\end{equation}
\par
We refer to \cite{balenascience,tesipalmisano} for a discussion
on the possibility to characterize (\ref{eq:back cov}) as a legitimate
implementation of the relativity principle.


\subsection{Unavailability of Hopf composition laws in the anti-de Sitter case}
Our next task is to find the scalars $\omega_{n}$ in (\ref{eq:generalLorentzW}), compatible
with (\ref{eq:Ortho Constraints}) and such that the overall composition law is associative. A solution can be found for $v_{\lambda}=\delta_{\lambda}^{1}$ in the form:
\begin{equation}
\left(q\oplus_{H}k\right)_{\lambda}=\frac{1}{\kappa}\left(k\cdot v+k_{4}\right)q_{\lambda}+k_{\lambda}+v_{\lambda}\frac{1}{\left(q\cdot v+q_{4}\right)}\left(-q\cdot k+\left(\kappa-q_{4}\right)k\cdot v\right)\label{eq:kdS with stabilizer}
\end{equation}
and making $v$ explicit, we have:
\begin{equation}
\left(q\oplus_{H}k\right)_{\lambda}=\frac{1}{\kappa}\left(k_{1}+k_{4}\right)q_{\lambda}+k_{\lambda}+\delta_{\lambda}^{1}\frac{1}{\left(q_{1}+q_{4}\right)}\left(-q\cdot k+\left(\kappa-q_{4}\right)k_{1}\right)\, .\label{eq:kdS with stabylizer-1}
\end{equation}
In the appendix \ref{appendix_anti2} we provide the proof of the uniqueness of the solution
(\ref{eq:kdS with stabylizer-1}) up to second order in the Planck
length. Assuming that also in the exact case the solution is unique,
let us analyze its structure. The point of interest for us is the
fact that the above composition law is not defined for $q_{1}+q_{4}=0$.
A way out of this problem could be found by performing a change of
coordinates such that the point $q_{1}+q_{4}$ is sent to infinity.
An example of this kind are the so called Horospherical coordinates
defined in the sector $p_{1}+p_{4}>0$ (for simplicity we work in
$1+1$ dimensions, but the result is valid for any dimensions) by:
\begin{equation}
\begin{cases}
\tilde{p}_{1}=\kappa\ln\left[\frac{1}{\kappa}\left(p_{1}+p_{4}\right)\right]=f_{1}\left(p\right),\\
\tilde{p}_{0}=\frac{\kappa p_{0}}{p_{1}+p_{4}}=f_{0}\left(p\right),\\
p_{1}=\kappa\sinh\left(\frac{1}{\kappa}\tilde{p}_{1}\right)+\frac{1}{2\kappa}\tilde{p}_{0}^{2}e^{\tilde{p}_{1}/\kappa}=\bar{f}_{1}\left(\tilde{p}\right),\\
p_{0}=e^{\tilde{p}_{1}/\kappa}\tilde{p}_{0}=\bar{f}_{0}\left(\tilde{p}\right),\\
p_{4}=\kappa \cosh\left(\frac{1}{\kappa}\tilde{p}_{1}\right)+\frac{1}{2\kappa}\tilde{p}_{0}^{2}e^{\tilde{p}_{1}/\kappa}=\digamma\left(\tilde{p}\right).
\end{cases}\label{eq:Horospherical map}
\end{equation}
In the Horospherical coordinates the composition laws indeed are free
from singularities and take the form:\footnote{We use here the same symbol $\oplus_{H}$ for the composition law
both in embedding and in comoving coordinates in order to simplify
the notation.}
\begin{equation}
\begin{cases}
\left(\tilde{q}\oplus_{H}\tilde{k}\right)_{0}=\tilde{q}_{0}+\tilde{k}_{0}e^{-\tilde{q}_{1}/\kappa}\, ,\\
\left(\tilde{q}\oplus_{H}\tilde{k}\right)_{1}=\tilde{q}_{1}+\tilde{k}_{1}\, .
\end{cases}\label{eq:horo comp}
\end{equation}

The problem here arises when we map the (standard) Lorentz transformations
which preserve the on-shell relation (\ref{eq:kappaAdSdispersion-1})
to horospherical coordinates. We find in fact that:
\begin{align}
\tilde{\Lambda}\left(\xi,\tilde{p}\right)=\begin{cases}
\tilde{\Lambda}{}_{0}\left(\tilde{p}\right)=&\kappa\frac{\cosh\left(\xi\right)e^{\tilde{p}_{1}/\kappa}\tilde{p}_{0}+\sinh\left(\xi\right)\left[\kappa\sinh\left(\tilde{p}_{1}/\kappa\right)+\frac{1}{2\kappa}\tilde{p}_{0}^{2}e^{\tilde{p}_1/\kappa}\right]}{\sinh\left(\xi\right)e^{\tilde{p}_{1}/\kappa}\tilde{p}_{0}+\cosh\left(\xi\right)\left[\kappa\sinh\left(\tilde{p}_{1}/\kappa\right)+\frac{1}{2\kappa}\tilde{p}_{0}^{2}e^{\tilde{p}_{1}/\kappa}\right]+\kappa\cosh\left(\tilde{p}_{1}/\kappa\right)-\frac{1}{2\kappa}\tilde{p}_{0}^{2}e^{\tilde{p}_{1}/\kappa}}\, ,\\
\tilde{\Lambda}{}_{1}\left(\tilde{p}\right)=&\kappa\ln\left\{\frac{1}{\kappa}\left[\cosh\left(\xi\right)\left[\kappa\sinh\left(\tilde{p}_{1}/\kappa\right)+\frac{1}{2\kappa}\tilde{p}_{0}^{2}e^{\tilde{p}_{1}/\kappa}\right]+\sinh\left(\xi\right)e^{\tilde{p}_{1}/\kappa}\tilde{p}_{0}+\right.\right.\\
&\left.\left.\kappa\cosh\left(\tilde{p}_{1}/\kappa\right)-\frac{1}{2\kappa}\tilde{p}_{0}^{2}e^{\tilde{p}_{1}/\kappa}\right]\right\}\, ,
\end{cases}
\end{align}
and these are ill-defined for:
\begin{equation}
\cosh\left(\xi\right)\bar{f}_{1}\left(\tilde{p}\right)+\sinh\left(\xi\right)\bar{f}_{0}\left(\tilde{p}\right)+F\left(\tilde{p}\right)=0\, ,
\end{equation}
which is always solvable.

The main point is that the map (\ref{eq:Horospherical map}) is ill-defined
for $p_{1}+p_{4}=0$ and that this constraint breaks evidently the
covariance of the theory. A similar observation was also done in \cite{Arzano:2014jua}. This is a general result in that in order
to avoid the singularity in $p_{1}+p_{4}$ we have to use a map that
is singular in $p_{1}+p_{4}$.

We therefore conclude that it is not possible, when adopting
the $3+1$-dimensional anti-de Sitter metric, to find a composition law
satisfying the Hopf properties (\ref{eq:IdentityExistence})-(\ref{eq:covariance}).


\subsection{Hopf composition laws for the  de Sitter case}\label{deSitterHopfcomp}
In relation to the obstruction we uncovered for the anti-de Sitter case,
one might gain some insight by looking at the composition law
that is found for the de Sitter case
by enforcing isometricity and associativity:
\begin{equation}
\left(q\oplus_{H}k\right)_{\lambda}=\frac{1}{\kappa}\left(k_{0}+k_{4}\right)q_{\lambda}+k_{\lambda}+\delta_{\lambda}^{0}\frac{1}{\left(q_{0}+q_{4}\right)}\left(-q\cdot k+\left(\kappa-q_{4}\right)k_{0}\right).\label{eq:dScomp with explicit stabilizer}
\end{equation}
This composition law is evidently ill defined for $q_{0}+q_{4}=0$,
but in the de Sitter case it is possible to enforce the requirement
 $q_{0}+q_{4}>0$ as a relativistic feature, since $q_{0}+q_{4}>0$
  is
invariant under the action of the Lorentz boosts on the de Sitter momentum space
and is also preserved
by the composition law, meaning that if $q_{0}+q_{4}>0$ and $k_{0}+k_{4}>0$,
than we will also have $\left(q\oplus_{H}k\right)_{0}+\left(q\oplus_{H}k\right)_{4}>0$.
Moreover, there is no obstruction in setting $q_{4}>0$ and
 $q_{0}>0$, so that $q_{0}+q_{4}>0$ is guaranteed. On the contrary,
in the anti-de Sitter case the condition $q_{1}>0$ would introduce,
severe interpretational challenges ($q_{0}>0$ is for positive energies,
which is physically desireable,
while $q_{1}>0$ would suggest positive spatial momentum in the $1$ direction,
the pathological requirement of motion only in the positive direction of a spatial axis).

We therefore have a good picture ($q_{4}>0$ and
 $q_{0}>0$) such that in the de Sitter case the composition law is nonsingular. Still
 it is interesting to notice that in the de Sitter
case the removal of the divergence for $q_{0}+q_{4}=0$ has different outcome with respect
to the anti-de Sitter case. By adopting comoving coordinates defined (in the portion
of de Sitter $p_{0}+p_{4}>0$) by:
\begin{equation}
\begin{cases}
\tilde{p}_{0}=\kappa\ln\left[\frac{1}{\kappa}\left(p_{0}+p_{4}\right)\right]=f_{0}\left(p\right),\\
\tilde{p}_{1}=\frac{\kappa p_{1}}{p_{0}+p_{4}}=f_{1}\left(p\right),\\
p_{0}=\kappa \sinh\left[\frac{1}{\kappa}\tilde{p}_{0}\right]+\frac{1}{2\kappa}\tilde{p}_{1}^{2}e^{\tilde{p}_{0}/\kappa}=\bar{f}_{0}\left(\tilde{p}\right),\\
p_{1}=e^{\tilde{p}_{0}/\kappa}p_{1}=\bar{f}_{1}\left(\tilde{p}\right),\\
p_{4}=\kappa \cosh\left(\frac{1}{\kappa}\tilde{p}_{0}\right)+\frac{1}{2\kappa}\tilde{p}_{1}^{2}e^{\tilde{p}_{0}/\kappa}=\digamma\left(\tilde{p}\right),
\end{cases}
\end{equation}
the composition law (\ref{eq:dScomp with explicit stabilizer}) takes
the form:
\begin{equation}
\begin{cases}
\left(\tilde{q}\oplus_{H}\tilde{k}\right)_{0}=\tilde{q}_{0}+\tilde{k}_{0}\, ,\\
\left(\tilde{q}\oplus_{H}\tilde{k}\right)_{1}=\tilde{q}_{1}+\tilde{k}_{1}e^{-\tilde{q}_{0}/\kappa}\, ,
\end{cases}
\end{equation}
which is clearly well-defined and corresponds to the $\kappa$-Poincar\'e
composition law discussed in subsection \ref{sub:-Poincar=0000E9-Momentum-Space}.
In particular the Lorentz boost in comoving coordinates takes the form
(\ref{eq:kappaboost}) that here we report again:
\begin{align}
\tilde{\Lambda}\left(\xi,\tilde{p}\right)=\begin{cases}
\tilde{\Lambda}{}_{0}\left(\tilde{p}\right)=&\kappa\ln\left\{\frac{1}{\kappa}\left[\cosh\left(\xi\right)\left[\kappa\sinh\left(\tilde{p}_{0}/\kappa\right)+\frac{1}{2\kappa}\tilde{p}_{1}^{2}e^{\tilde{p}_{0}/\kappa}\right]+\sinh\left(\xi\right)e^{\tilde{p}_{0}/\kappa}\tilde{p}_{1}\right.\right.\\
&\left.\left.+\kappa\cosh\left(\tilde{p}_{0}/\kappa\right)-\frac{1}{2\kappa}\tilde{p}_{1}^{2}e^{\tilde{p}_{0}/\kappa}\right]\right\}\, ,\\
\tilde{\Lambda}{}_{1}\left(\tilde{p}\right)=&\kappa\frac{\cosh\left(\xi\right)e^{\tilde{p}_{0}/\kappa}\tilde{p}_{1}+\sinh\left(\xi\right)\left[\kappa\sinh\left(\tilde{p}_{0}/\kappa\right)+\frac{1}{2\kappa}\tilde{p}_{1}^{2}e^{\tilde{p}_0/\kappa}\right]}{\sinh\left(\xi\right)e^{\tilde{p}_{0}/\kappa}\tilde{p}_{1}+\cosh\left(\xi\right)\left[\kappa\sinh\left(\tilde{p}_{0}/\kappa\right)+\frac{1}{2\kappa}\tilde{p}_{1}^{2}e^{\tilde{p}_{0}/\kappa}\right]+\kappa\cosh\left(\tilde{p}_{0}/\kappa\right)-\frac{1}{2\kappa}\tilde{p}_{1}^{2}e^{\tilde{p}_{0}/\kappa}}\, ,
\end{cases}\label{eq:de sitter boost 2}
\end{align}
\par
It is easy to check that, unlike the analogous anti-de Sitter case,
the transformation (\ref{eq:de sitter boost 2}) is well-defined in
the sector $p_{0}>0$.\footnote{The boost (\ref{eq:de sitter boost 2}) would be ill defined for $\cosh\left(\xi\right)\bar{f}_{0}\left(\tilde{p}\right)
+\sinh\left(\xi\right)\bar{f}_{1}\left(\tilde{p}\right)+\digamma\left(\tilde{p}\right)=0$,
but this condition is not solvable for a positive $\tilde{p}_{0}$
(or, which is the same, a positive $p_{0}$).}

\section{Alternative Scenarios with anti-de Sitter momentum space\label{sec:Alternative-Scenarios-with}}

In the previous section we have seen that in the realm of DSR-compatible
(relativistic) deformations it is not possible
to require the properties (\ref{eq:IdentityExistence}), (\ref{eq:Isometricity}), (\ref{eq:Associativity}), (\ref{eq:covariance})
({\it i.e.} $R_1$, $R_2$, $R_3$, $R_4$)
for composition laws implemented in $3+1$-dimensional anti-de
Sitter momentum space. Our next task is consider the possibility
of composition laws which are explicitly relativistically covariant but are such that
either  the condition
of associativity or the  condition of isometricity are lost.

\subsection{Explicit covariance and isometricity: anti-Snyder momentum space}

As we have seen in section \ref{sub:Explicitelly-Covariant-Isometric},
the two possible composition laws that satisfy both the conditions
of explicit covariance and isometricity, at the expense of the associativity
condition, are those of Eq. (\ref{eq:AntiSnyder comp laws}), that we
here report:
\begin{equation}
\begin{cases}
1) & \begin{cases}
\left(p\oplus_{(1)}q\right)_{\lambda}=\frac{1}{\kappa}p_{\lambda}q_{4}+\left(\delta_{\lambda}^{\alpha}-\frac{p_{\lambda}p^{\alpha}}{\kappa(\kappa+p_{4})}\right)q_{\alpha}\, ,\\
\left(p\oplus_{(1)}q\right)_{4}=\frac{1}{\kappa}\left(p_{4}q_{4}-p^{\alpha}q_{\alpha}\right).
\end{cases}\\
2) & \begin{cases}
\left(p\oplus_{(2)}q\right)_{\lambda}=\frac{1}{\kappa}p_{\lambda}q_{4}+\left(\delta_{\lambda}^{\alpha}-\frac{p_{\lambda}p^{\alpha}}{\kappa(\kappa-p_{4})}\right)q_{\alpha}\, ,\\
\left(p\oplus_{(2)}q\right)_{4}=\frac{1}{\kappa}\left(p_{4}q_{4}+p^{\alpha}q_{\alpha}\right).
\end{cases}
\end{cases}\label{eq:AntiSnyder comp laws-1}
\end{equation}
\par
We can see that both the composition laws are divergent respectively
for $\kappa+p_{4}=0$ and $\kappa-p_{4}=0$. Furthermore, it is not
possible to choose a portion of anti-de Sitter such that $\kappa+p_{4}>0$
and close under the action of $\oplus_{(1)}$ and the Lorentz boosts,
nor any portion characterized by $\kappa-p_{4}=0$ and close under
the action of $\oplus_{(2)}$ and the Lorentz boosts.\footnote{Considering for example the first composition law $\oplus_{(1)}$,
this can be seen by fixing, for any couple of momenta $q$ and $p$,
the values of $q_{4}$ and $p_{4}$ , and of the on-shell relations
$m_{q}^{2}=q_{0}^{2}-\vec{q}^{2}$ and $m_{p}^{2}=p_{0}^{2}-\vec{p}^{2}$
. By picking the important case in which the particle $q$ is at rest,
then the condition $p_{4}q_{4}-p^{\alpha}q_{\alpha}$ translates in
the constraint $\left(\kappa^{2}-m_{p}^{2}\right)\frac{q_{4}^{2}}{m_{q}^{2}}-m_{p}^{2}>p_{1}^{2}$,
which is not admissible because is not preserved by the action of
the Lorentz group.}
\par
Although the two composition laws alone seem to have some pathologies,
still it is possible to construct a well-defined composition law from
(\ref{eq:AntiSnyder comp laws-1}) with this prescription:
\begin{equation}
\left(p\oplus_{\kappa}q\right)=\begin{cases}
\left(p\oplus_{(1)}q\right)\ ;\  & p_{4}\geq0\, ,\\
\left(p\oplus_{(2)}q\right)\ ;\  & p_{4}<0\, .
\end{cases}\label{eq:welldefinedAntiSnyder}
\end{equation}
\par
Because of the resemblance of this composition law with the Snyder
one discussed in subsection \ref{sub:Proper-and-Snyder}, we will
refer in the following to (\ref{eq:welldefinedAntiSnyder}) as the
anti-Snyder composition law.


\subsection{Explicit covariance and associativity}

We consider now the case of an explicitly covariant and associative
(but non-isometric) composition law. The general explicitly covariant
composition law can be written as:
\begin{equation}
\left(q\oplus_{\kappa}k\right)_{\lambda}=A\left(q,k\right)q_{\lambda}+B\left(q,k\right)k_{\lambda}\, ,
\end{equation}
where $A$ and $B$ are scalars, {\it i. e.} arbitrary functions of $q^{2}$,
$k^{2}$ and $q\cdot k$.

Imposing associativity we get the system of equations:
\begin{equation}
\begin{cases}
A\left(\left(q\oplus k\right),p\right)A\left(q,k\right)=A\left(q,\left(k\oplus p\right)\right),\\
A\left(\left(q\oplus k\right),p\right)B\left(q,k\right)=B\left(q,\left(k\oplus p\right)\right)A\left(k,p\right),\\
B\left(\left(q\oplus k\right),p\right)=B\left(q,\left(k\oplus p\right)\right)B\left(k,p\right).
\end{cases}\label{eq:Associativity Conditions}
\end{equation}
\par
We search for a perturbative solution of (\ref{eq:Associativity Conditions})
at second order in $\kappa^{-1}$. We then expand the scalars
$A$ and $B$ as follows:
\begin{equation}
\begin{cases}
A\left(q,k\right)=1+\frac{1}{\kappa^{2}}\left(bk^{2}+cq\cdot k\right),\\
B\left(q,k\right)=1+\frac{1}{\kappa^{2}}\left(dq^{2}+eq\cdot k\right),
\end{cases}
\end{equation}
where $b$, $c$, $d$ and $e$ are constants. Inserting the expansion
above in (\ref{eq:Associativity Conditions}) we get:
\begin{equation}
\begin{cases}
\left[1+\frac{1}{\kappa^{2}}\left(bp^{2}+c\left(q+k\right)\cdot p\right)\right]\left[1+\frac{1}{\kappa^{2}}\left(bk^{2}+cq\cdot k\right)\right]=\left[1+\frac{1}{\kappa^{2}}\left(b\left(k+p\right)^{2}+cq\cdot\left(k+p\right)\right)\right]\, ,\\
\left[1+\frac{1}{\kappa^2}\left(bp^{2}+c\left(q+k\right)\cdot p\right)\right]\left[1+\frac{1}{\kappa^2}\left(dq^{2}+eq\cdot k\right)\right]=\left[1+\frac{1}{\kappa^{2}}\left(dq^{2}+eq\cdot\left(k+p\right)\right)\right]\left[1+\frac{1}{\kappa^{2}}\left(bp^{2}+cp\cdot k\right)\right]\, ,\\
\left[1+\frac{1}{\kappa^2}\left(d\left(q+k\right)^{2}+e\left(q+k\right)\cdot p\right)\right]=\left[1+\frac{1}{\kappa^2}\left(dq^{2}+eq\cdot\left(k+p\right)\right)\right]\left[1+\frac{1}{\kappa^{2}}\left(dk^{2}+ep\cdot k\right)\right]\, ,
\end{cases}
\end{equation}
which is solved by:
\begin{equation}
\begin{cases}
c=2b,\\
c=e,\\
2d=e,
\end{cases}
\end{equation}
and we get the unique solution:
\begin{equation}
\left(q\oplus_{\kappa}k\right)_{\lambda}=q_{\lambda}+k_{\lambda}+\frac{1}{\kappa^{2}}\left[b\left(k^{2}+2q\cdot k\right)q_{\lambda}+b\left(q^{2}+2q\cdot k\right)k_{\lambda}\right]\, .
\end{equation}


\subsection{Explicit covariance and commutativity: Proper anti-de Sitter}

The last possibility that we will consider is an example of a composition
law which is neither associative nor isometric, but is commutative. We proceed
 in analogy with the proper de Sitter
composition law discussed in subsection \ref{sub:Proper-and-Snyder}, so we shall label
this option as the ``Proper anti-de Sitter'' composition law.

Adopting Cartesian parameterization (see the appendix \ref{appendix_anti1}) and substituting
in (\ref{eq:polyakovsecondorder}) the anti-de Sitter Levi-Civita
connection $\Gamma_{\lambda}^{\mu\nu}=\frac{1}{\kappa^{2}}p_{\lambda}g^{\mu\nu}\left(p\right)$
we then get:
\begin{equation}
\left(q\oplus_{\kappa}p\right){}_{\lambda}=\left(q_{\lambda}+k_{\lambda}\right)\left(1-\frac{1}{2\kappa^{2}}q\cdot k\right).
\end{equation}
\par
We notice that the Proper anti-de Sitter composition law is non-associative
but commutative and covariant under the action of the standard Lorentz
transformations.


\section{Aside on the Magueijo-Smolin proposal}\label{sec:Magueijo-Smolin-Composition}
The notion of DSR-relativistic theory was first introduced by one of us
in the studies \cite{Amelinodsr1,Amelinodsr2}, using as illustrative model
a description of relativistic symmetries which within the relative-locality framework
should be formulated in terms of a de Sitter momentum space.
A second DSR-relativistic model was then introduced in a follow-up study by Magueijo
and Smolin, and we shall here show that this Magueijo-Smolin model
can be formulated in terms of anti-de Sitter momentum space.
This is interesting since  in a previous study \cite{Granik:2002wd}
the Magueijo-Smolin model was erroneously described in terms of the bicrossproduct basis of $\kappa$-Poincar\'e,
which would imply a description in terms of a de Sitter momentum space. We shall explain the origin of that confusion and
also analyze the Magueijo-Smolin model in light of our general
results concerning theories with anti-de Sitter momentum space.

The Magueijo-Smolin model was inspired by some work by
Fock \cite{fock1969theory} for spacetime, basically adopting
a momentum-space picture which reproduces the Fock spacetime picture.
Central to the Magueijo-Smolin proposal is the map $U(p_{0},\kappa)$, such that
\begin{align}
p_{\mu}\mapsto U(p_{0},\kappa)\triangleright p_{\mu},\\
U(p_{0},\kappa)\triangleright p_{\mu}=\frac{p_{\mu}}{1-p_{0}/\kappa}.\nonumber
\end{align}
This map induces a deformation of boost transformations and a dispersion relation given by
\begin{equation}
\|p\|^{2}=\frac{\eta^{\mu\nu}p_{\mu}p_{\nu}}{(1-p_{0}/\kappa)^{2}}=\eta^{\mu\nu}(U(p_{0},\kappa)\triangleright p_{\mu})(U(p_{0},\kappa)\triangleright p_{\nu}).
\end{equation}

In order to provide a geometric characterization of the Magueijo-Smolin proposal
let us choose the following set of coordinates for anti-de Sitter momentum space:
\begin{eqnarray}
P_{\mu}=\frac{\kappa p_{\mu}}{\sqrt{p^{2}+(\kappa-p_{0})^{2}}},\label{mscoordinates}\\
P_{4}=\frac{\kappa(\kappa-p_{0})}{\sqrt{p^{2}+(\kappa-p_{0})^{2}}},\label{mscoordinates1}
\end{eqnarray}
where $p^{2}\equiv p_{0}^{2}-p_{1}^{2}-p_{2}^{2}-p_{3}^{2}$.

We rely again on the relative-locality framework
to describe the on-shell relation in terms
of the geodesic distance from the origin of momentum space,
intended as $(0,0,0,0,\kappa)$, to an arbitrary point in
the manifold. This gives $P_{4}^{2}=\kappa^{2}\cos^{2}(d/\kappa)$,
which can be rewritten as follows:
\begin{equation}
P_{4}^{2}=\kappa^{2}\cos^{2}(d/\kappa)
=\frac{\kappa^{2}(\kappa-p_{0})^{2}}{p^{2}
+(\kappa-p_{0})^{2}}\Rightarrow\|p\|^{2}:=\frac{p^{2}}{(1-p_{0}/\kappa)^{2}}=\kappa^{2}\tan^{2}(d/\kappa).
\end{equation}

Therefore the Magueijo-Smolin on-shellness relation
is a function of the geodesic distance in anti-de Sitter momentum space.
 For a de Sitter momentum space, which we can obtain by performing a Wick rotation,
one finds a functional dependence on a hyperbolic tangent function, rather than
a tangent function. The hyperbolic tangent is a bounded function, leading to
a description of the mass of particles which has a maximum allowed value,
whereas Magueijo and Smolin specified very explicitly \cite{2002PhRvL..88s0403M}
that they were proposing a picture with unbounded mass.
From this one must inevitably conclude that the geometric picture of the
Magueijo-Smolin proposal must be given in terms of a momentum space with
anti-de Sitter geometry.


The work of Magueijo and Smolin on DSR-relativistic scenarios also gives us
another structure that deserves to be analyzed from our perspective.
This is the proposal for the law of composition of momenta
which they put forward in Ref.\cite{2003PhRvD..67d4017M},
a proposal that was intended to ensure that the deformation
of relativistic laws would not affect significantly  macroscopic systems (composed of
many fundamental particles). In that proposal
the total momentum, $p_{\mu}^{(1,...,N)} \equiv p^{(1)}\oplus...\oplus p^{(N)}$, obtained composing $N$
fundamental-particle momenta, is described implicitly
through
\begin{equation}
U(p_{0}^{(1,...,N)},N\kappa)\triangleright(p^{(1)}\oplus...\oplus p^{(N)})=U(p_{0}^{(1)},\kappa)\triangleright p^{(1)}+...+U(p_{0}^{(N)},\kappa)\triangleright p^{(N)} \, .
\end{equation}
When written in terms of the coordinates that we are using, this definition
takes the form:
\begin{equation}
\frac{p_{\mu}^{(1,...,N)}}{1-p_{0}^{(1,...,N)}/N\kappa}=\sum_{i}^{N}\frac{p_{\mu}^{(i)}}{1-p_{0}^{(i)}/\kappa},\label{mscomplaw}
\end{equation}
For this composed momentum $p_{\mu}^{(1,...,N)}$ one finds
that it is governed by Lorentz transformations of the same form as
the fundamental-particle Lorentz transformations,
but with deformation parameter rescaled from $\kappa$
to $N\kappa$.
Of course this also implies that the on-shelleness condition for the composed  momentum
has the same form as the on-shelleness condition for fundamental particles,
but with deformation parameter rescaled from $\kappa$
to $N\kappa$.

We can explicitly give a geometrical interpretation for this kind
of composition law. Let $\mathbb{M}^{5}$ be the $5$-dimensional
Minkowski space with signature $(+---+)$, labeled by Cartesian coordinates
$\{P_{A}\}$, $A=0,...,4$, let us denote with $adS_{\kappa}$, the anti-de Sitter
space with curvature $\kappa^{-1}$, the subset of $\mathbb{M}^{5}$
satisfying the constraint $(P_{0})^{2}-(P_{1})^{2}-(P_{2})^{2}-(P_{3})^{2}+(P_{4})^{2}=\kappa^{2}$,
and let ${\cal P}\subset adS_{\kappa}=\{P^{(i)}\}=\{P_{A}^{(i)}\}$, $i=1,...,N$,
be the set of $N$ points defined in $adS_{\kappa}$.
We can then introduce a composition law $\bigoplus_{i=1}^{N}$ as the map
\begin{align}
\bigoplus_{i=1}^{N}:{\cal P}\subset adS_{\kappa}\subset\mathbb{M}^{5}\longrightarrow adS_{N\kappa}\subset\mathbb{M}^{5},\nonumber \\
P_{A}^{(i)}\mapsto\bigoplus_{i=1}^{N}P_{A}^{(i)}=P_{A}^{(1,...,N)},{}A=0,...,4,
\end{align}
where
\begin{align}
P_{\mu}^{(1,...,N)}=\kappa\left(\sum_{i=1}^{N}\frac{P_{\mu}^{(i)}}{P_{4}^{(i)}}\right)\left[1+\frac{1}{N^{2}}\eta^{\alpha\beta}\sum_{j,k=1}^{N}\left(\frac{P_{\alpha}^{(j)}}{P_{4}^{(j)}}\right)\left(\frac{P_{\beta}^{(k)}}{P_{4}^{(k)}}\right)\right]^{-1/2}{},{}\mu=0,...,3,\\
P_{4}^{(1,...,N)}=N\kappa\left[1+\frac{1}{N^{2}}\eta^{\mu\nu}\sum_{i,j=1}^{N}\left(\frac{P_{\mu}^{(i)}}{P_{4}^{(i)}}\right)\left(\frac{P_{\nu}^{(j)}}{P_{4}^{(j)}}\right)\right]^{-1/2}.
\end{align}

It is straightforward to verify that $\eta^{\mu\nu}P_{\mu}^{(1,...,N)}P_{\nu}^{(1,...,N)}+\left(P_{4}^{(1,...,N)}\right)^{2}=(N\kappa)^{2}$.
For the proposal of Magueijo and Smolin \cite{2002PhRvL..88s0403M} we have
\begin{equation}
\frac{p_{\mu}^{(1,...,N)}}{1-p_{0}^{(1,...,N)}/N\kappa}=\sum_{i=1}^{N}\frac{p_{\mu}^{(i)}}{1-p_{0}^{(i)}/\kappa}.\label{defcomplaw}
\end{equation}

Form our perspective it noteworthy that this
Magueijo-Smolin proposal for the law of composition of momenta  is commutative but not associative, and is geometrically described by a map between anti-de Sitter spaces ({\it i.e.}, from a anti-de Sitter
space with curvature given by the scale $\kappa^{-1}$ to another anti-de Sitter space with
curvature $(N\kappa)^{-1}$).

\section{Concluding remarks}
Our analysis intends to affect the future development of research inspired by recent results
supporting the possibility that momentum space is curved in several approaches to the quantum-gravity
problem.
While in the 4D context different approaches provide different expectations, in 3D quantum gravity
all approaches agree on the fact that momentum space has anti-de Sitter geometry, with certain
characterizing properties which we here highlighted, $R1$, $R2$, $R3$, $R4b$. It would be
natural then to explore the possibility that also 4D quantum gravity
might be characterized by an anti-de Sitter momentum space
with those properties.
To our surprise our investigations showed that this is not possible: in a 4D context
with anti-de Sitter momentum space the properties $R1$, $R2$, $R3$ are incompatible with
the type of relativistic covariance coded in $R4$ and are also incompatible with the
weaker notion of relativistic covariance coded in our requirement $R4b$.

Those attempting to use the results on 3D quantum gravity as guidance toward
4D quantum gravity should take our findings into account.
The analogy between 3D quantum gravity and 4D quantum gravity cannot be too close,
at least for what concerns the aspects on which we here focused.
Our perspective is that 3D gravity and 4D gravity clearly have some important differences, already at
the classical level (3D gravity has no propagating gravitational degrees of freedom), and
the analogy, if at all fruitful, can only be partial.
In the final part of our study we 
investigated the possibility of 4D momentum spaces with only some of the properties
$R1$, $R2$, $R3$ and $R4$/$R4b$, and we feel that a broader investigation of this possibility might be
valuable.

\subsection*{Acknowledgements}
G.A.-C. acknowledges financial support by the Programme STAR Plus, funded by Federico II University and Compagnia di San Paolo, and by the MIUR, PRIN 2017 grant 20179ZF5KS. IPL was partially supported by the National Council for Scientific and Technological Development - CNPq grant 306414/2020-1 and by the Grant 3197/2021, Paraiba State Research Foundation (FAPESQ). The authors would like to acknowledge networking support by the COST Action QGMM (CA18108), supported by COST (European Cooperation in Science and Technology).
\appendix

\section{Cartesian, Comoving and Horospherical Coordinates}\label{appendix_anti1}
\subsection{De Sitter space}
A useful setting one can choose to analyse de Sitter space consists in embedding it in a flat, higher-dimensional manifold \cite{2016IJMPS..4160126L,moschella2006sitter}. We use Greek letters to label coordinates for $0,...,3$. One can describe a $4$-dimensional de Sitter manifold as a hypersurface embedded in a $5$-dimensional Minkowski space with signature $(+----)$, with an extra space-like coordinate. For a Minkowski metric
\begin{equation}\label{appendix_cartesian1}
ds^2=dp_0^2-dp_1^2-dp_2^2-dp_3^2-dp_4^2,
\end{equation}
de Sitter space is defined as the hyperboloid satisfying
\begin{equation}
p_0^2-p_1^2-p_2^2-p_3^2-p_4^2\doteq p^2-p_4^2=-\kappa^2.
\end{equation}


\subsubsection{Cartesian coordinates for de Sitter}
Another possible choice of coordinate system for de Sitter space is simply to use the coordinates of the embedding and pick half of de Sitter space, for instance the one with
\begin{equation}
p_4=\sqrt{p^2+\kappa^2}>0.
\end{equation}
\par
This way, by calculating the differential $dp_4$ and substituting in \ref{appendix_cartesian1}, we find the line element
\begin{equation}
ds^2=g^{\mu\nu}dp_{\mu}dp_{\nu},
\end{equation}
for
\begin{equation}
g^{\mu\nu}=\eta^{\mu\nu}-\frac{p^{\mu}p^{\nu}}{p^2+\kappa^2}\doteq\eta^{\mu\nu}-\frac{p^{\mu}p^{\nu}}{p_4^2}.
\end{equation}


\subsubsection{Comoving coordinates}
\par
Another possible choice of coordinate system for de Sitter space is
\begin{align}
\left\{
\begin{array}{rcl}
p_0&=&\kappa \sinh(\tilde{p}_0/\kappa)+e^{\tilde{p}_0/\kappa}(\tilde{p}_1^2+\tilde{p}_2^2+\tilde{p}_3^2)/2\kappa,\\
p_i&=&e^{\tilde{p}_0/\kappa} \tilde{p}_i,\\
p_4&=&\kappa\cosh(\tilde{p}_0/\kappa)-e^{\tilde{p}_0/\kappa}(\tilde{p}_1^2+\tilde{p}_2^2+\tilde{p}_3^2)/2\kappa,
\end{array}\label{comoving}
\right.
\end{align}
\par
These are called, comoving coordinates. The metric assumes the following form:
\begin{equation}
ds^2=d\tilde{p}_0^2-e^{2\tilde{p}_0/\kappa}\left(d\tilde{p}_1^2+d\tilde{p}_2^2+d\tilde{p}_3^2\right),
\end{equation}
This is the coordinate system usually used to describe $\kappa$-Poincar\'e momentum space \cite{GiuliaFlavio}. It covers only half of de Sitter space, the portion $p_0+p_4>0$.


\subsection{Anti-de Sitter space}
The anti-de Sitter metric can be found using a similar procedure of embedding a hypersurface in a flat space \cite{2016IJMPS..4160126L,moschella2006sitter}. Let us do the example of embedding this $4$-dimensional manifold in a $5$-dimensional one with signature $(+---+)$. For the Minkowski metric
\begin{equation}
ds^2=dp_0^2-dp_1^2-dp_2^2-dp_3^2+dp_4^2,
\end{equation}
anti-de Sitter space is defined as the hyperboloid satisfying
\begin{equation}
p_0^2-p_1^2-p_2^2-p_3^2+P_4^2\doteq p^2+p_4^2=\kappa^2.
\end{equation}


\subsubsection{Cartesian coordinates for anti-de Sitter}
Following a similar procedure of the last section, we might pick half of anti-de Sitter space, for instance
\begin{equation}
p_4=\sqrt{\kappa^2-p^2},
\end{equation}
to have
\begin{equation}
g^{\mu\nu}=\eta^{\mu\nu}+\frac{p^{\mu}p^{\nu}}{\kappa^2-p^2}\doteq \eta^{\mu\nu}+\frac{p^{\mu}p^{\nu}}{p_4^2},
\end{equation}
which can be derived from the de Sitter case by a Wick rotation $(p_4,\kappa)\mapsto (ip_4,i\kappa)$.


\subsubsection{Horospherical coordinates}
Another set of coordinates in similar to the comoving ones of de Sitter space and are called horospherical coordinates. They are defined by the embedding
\begin{align}
\left\{
\begin{array}{rcl}
p_0&=&\tilde{p}_0e^{\tilde{p}_1/\kappa},\\
p_1&=&\kappa\sinh\left(\tilde{p}_1/\kappa\right)-e^{\tilde{p}_1/\kappa}(-\tilde{p}_0^2+\tilde{p}_2^2+\tilde{p}_3^2)/2\kappa,\\
p_2&=&\tilde{p}_2e^{\tilde{p}_1/\kappa},\\
p_3&=&\tilde{p}_3e^{\tilde{p}_1/\kappa},\\
p_4&=&\kappa\cosh\left(\tilde{p}_1/\kappa\right)+e^{\tilde{p}_1/\kappa}(-\tilde{p}_0^2+\tilde{p}_2^2+\tilde{p}_3^2)/2\kappa,
\end{array}
\right.
\end{align}
and the line element assumes the form
\begin{equation}
ds^2=e^{2p_1/\kappa}\left(dp_0^2-dp_2^2-dp_3^2\right)-dp_1^2.
\end{equation}

\section{Uniqueness of Hopf Composition Law at Second Order in the Planck Lenght}\label{appendix_anti2}

In order to see that the composition law (\ref{eq:kdS with stabylizer-1}) is the unique solution
of (\ref{eq:IdentityExistence})-(\ref{eq:covariance}) up to an arbitrary rotation, we expand the composition law
(\ref{eq:generalkappaAdScomp-1}) and the general boost $\Omega$ at
second order in the Planck length:
\begin{equation}\label{unicity1}
\begin{cases}
\left(p\oplus_{H}q\right)_{\lambda}\approx p_{\lambda}+q_{\lambda}+\frac{1}{\kappa}\omega_{\lambda}^{\ \beta\gamma}q_{\beta}p_{\gamma}-\frac{1}{2}\frac{1}{\kappa^{2}}p_{\lambda}\left(q^{2}+q\cdot p\right)+\frac{1}{\kappa^{2}}X_{\lambda}^{\ \beta\gamma\lambda}p_{\gamma}p_{\lambda}q_{\beta}\, ,\\
\Omega_{\alpha}^{\ \beta}\approx\delta_{\alpha}^{\ \beta}+\frac{1}{\kappa}\omega_{\alpha}^{\ \beta\gamma}p_{\gamma}+\frac{1}{\kappa^{2}}X_{\alpha}^{\ \beta\gamma\lambda}p_{\gamma}p_{\lambda}\, ,
\end{cases}
\end{equation}
where the condition of (pseudo) orthogonality for $\Omega$
translates into the constraints:
\begin{equation}
\begin{cases}
\omega^{\mu\nu\gamma}+\omega^{\nu\mu\gamma}=0,\\
2X^{(\mu\nu)\alpha\beta}+\omega^{\theta\mu(\beta}\omega_{\theta}^{\ \nu\alpha)}=0.
\end{cases}\label{eq:Ortho Cond}
\end{equation}
\par
Imposing the condition of associativity for the composition law results
in:
\begin{equation}
0=\frac{1}{2}\left(\delta_{\lambda}^{\nu}\eta^{\alpha\mu}-\delta_{\lambda}^{\mu}\eta^{\alpha\nu}\right)+\omega_{\lambda}^{\ \alpha\theta}\omega_{\theta}^{\ \mu\nu}-\omega_{\lambda}^{\ \theta\nu}\omega_{\theta}^{\ \alpha\mu}+2X_{\lambda}^{\ \alpha\nu\mu}\, .\label{eq:associativity constraint}
\end{equation}
\par
Using the previous equations we find that $X$ is completely determined
as a function of $\omega$, which in turns has to satisfy the condition:
\begin{equation}
0=\frac{1}{2}\left(\delta_{\lambda}^{\nu}\eta^{\alpha\mu}-\delta_{\lambda}^{\mu}\eta^{\alpha\nu}\right)+\omega_{\lambda}^{\ \alpha\theta}\omega_{\theta}^{\ [\mu\nu]}-\omega_{\lambda}^{\ \theta[\nu}\omega_{\theta}^{\ \alpha\mu]}\, .\label{eq:omega constraint}
\end{equation}
\par
Parameterizing $\omega$ as:
\begin{equation}
\omega_{\lambda}^{\ \mu\nu}=c\epsilon_{\lambda}^{\ \mu\nu\delta}v_{\delta}+d\left(v_{\lambda}\eta^{\mu\nu}-\eta_{\lambda}^{\mu}v^{\nu}\right),\label{eq:par anisotr}
\end{equation}
and inserting (\ref{eq:par anisotr}) into (\ref{eq:omega constraint})
we get:
\begin{equation}
\delta_{\lambda}^{[\nu}\eta^{\alpha\mu]}+d^{2}v^{2}\delta_{\lambda}^{[\nu}\eta^{\alpha\mu]}=0,
\end{equation}
which tell us that $v$ has to be spacelike. In particular we will
choose $d^{2}=1=-v^{2}$ so that:
\begin{equation}\label{unicityw}
\omega_{\lambda}^{\ \mu\nu}=v_{\lambda}\eta^{\mu\nu}-\delta_{\lambda}^{\mu}v^{\nu}\, ,
\end{equation}
and the $X$ is given by (\ref{eq:associativity constraint}):
\begin{equation}\label{unicityX}
X_{\lambda}^{\ \alpha\nu\mu}=\frac{1}{2}\delta_{\lambda}^{(\nu}\eta^{\alpha\mu)}+v_{\lambda}v^{(\mu}\eta^{\alpha\nu)}-\frac{1}{2}v_{\lambda}\eta^{\mu\nu}v^{\alpha}\, ,
\end{equation}
and is unique up to a spatial (three dimensional) rotation of $w$.
\par
It is not hard to see that by expanding the composition law (\ref{eq:kdS with stabylizer-1}) at
second order in the Planck length we get indeed contribution of the
form (\ref{unicity1}), considering (\ref{unicityw}) and (\ref{unicityX}).

\bibliographystyle{utphys}
\bibliography{AdSBib3}

\providecommand{\href}[2]{#2}\begingroup\raggedright\begin{thebibliography}{}

\bibitem{Amelinodsr1}
G. Amelino-Camelia, ``{Relativity in space-times with short distance structure
  governed by an observer independent (Planckian) length scale},''
  \href{http://dx.doi.org/10.1142/S0218271802001330}{{\em Int. J. Mod. Phys.}
  {\bfseries D11} (2002) 35--60},
\href{http://arxiv.org/abs/gr-qc/0012051}{{\ttfamily arXiv:gr-qc/0012051
  [gr-qc]}}.

\bibitem{Amelinodsr2}
G. Amelino-Camelia, ``{Testable scenario for relativity with minimum length},''
  \href{http://dx.doi.org/10.1016/S0370-2693(01)00506-8}{{\em Phys. Lett.}
  {\bfseries B510} (2001) 255--263},
\href{http://arxiv.org/abs/hep-th/0012238}{{\ttfamily arXiv:hep-th/0012238
  [hep-th]}}.

\bibitem{Amelino-Camelia2013}
G.~Amelino-Camelia, ``{Quantum-Spacetime Phenomenology},''
  \href{http://dx.doi.org/10.12942/lrr-2013-5}{{\em Living Rev. Rel.}
  {\bfseries 16} (2013) 5},
\href{http://arxiv.org/abs/0806.0339}{{\ttfamily arXiv:0806.0339 [gr-qc]}}.

\bibitem{Addazi:2021xuf}
Addazi, A. and others, ``{Quantum gravity phenomenology at the dawn of the
  multi-messenger era\textemdash{}A review},''
  \href{http://dx.doi.org/10.1016/j.ppnp.2022.103948}{{\em Prog. Part. Nucl.
  Phys.} {\bfseries 125} (2022) 103948},
  \href{http://arxiv.org/abs/2111.05659}{{\ttfamily arXiv:2111.05659
  [hep-ph]}}.

\bibitem{freidellivineeffective}
L.~Freidel and E.~R.~Livine, ``{Effective 3-D quantum gravity and
  non-commutative quantum field theory},''
  \href{http://dx.doi.org/10.1103/PhysRevLett.96.221301}{{\em Phys. Rev. Lett.}
  {\bfseries 96} (2006) 221301},
\href{http://arxiv.org/abs/hep-th/0512113}{{\ttfamily arXiv:hep-th/0512113
  [hep-th]}}.

\bibitem{MatschullWelling}
H. J. Matschull and M. Welling, ``{Quantum mechanics of a point particle in
  (2+1)-dimensional gravity},''
  \href{http://dx.doi.org/10.1088/0264-9381/15/10/008}{{\em Class. Quant.
  Grav.} {\bfseries 15} (1998) 2981--3030},
\href{http://arxiv.org/abs/gr-qc/9708054}{{\ttfamily arXiv:gr-qc/9708054
  [gr-qc]}}.

\bibitem{RellocinAdS}
G.~Amelino-Camelia, M.~Arzano, S.~Bianco and R.~J.~Buonocore, ``{The
  DSR-deformed relativistic symmetries and the relative locality of 3D quantum
  gravity},'' \href{http://dx.doi.org/10.1088/0264-9381/30/6/065012}{{\em
  Class. Quant. Grav.} {\bfseries 30} (2013) 065012},
\href{http://arxiv.org/abs/1210.7834}{{\ttfamily arXiv:1210.7834 [hep-th]}}.

\bibitem{Amelino-Camelia:2016gfx}
G.~Amelino-Camelia, M.~M.~da Silva, M.~Ronco, L.~Cesarini and O.~M.~Lecian,
  ``{Spacetime-noncommutativity regime of Loop Quantum Gravity},''
  \href{http://dx.doi.org/10.1103/PhysRevD.95.024028}{{\em Phys. Rev.}
  {\bfseries D95} no.~2, (2017) 024028},
\href{http://arxiv.org/abs/1605.00497}{{\ttfamily arXiv:1605.00497 [gr-qc]}}.

\bibitem{Bojowald:2012ux}
M.~Bojowald and G.~M.~Paily, ``{Deformed General Relativity},''
  \href{http://dx.doi.org/10.1103/PhysRevD.87.044044}{{\em Phys. Rev.}
  {\bfseries D87} no.~4, (2013) 044044},
\href{http://arxiv.org/abs/1212.4773}{{\ttfamily arXiv:1212.4773 [gr-qc]}}.

\bibitem{Brahma:2016tsq}
S.~Brahma, M.~Ronco, G.~Amelino-Camelia and A.~Marciano, ``{Linking loop
  quantum gravity quantization ambiguities with phenomenology},''
  \href{http://dx.doi.org/10.1103/PhysRevD.95.044005}{{\em Phys. Rev.}
  {\bfseries D95} no.~4, (2017) 044005},
\href{http://arxiv.org/abs/1610.07865}{{\ttfamily arXiv:1610.07865 [gr-qc]}}.

\bibitem{Cianfrani:2016ogm}
F.~Cianfrani, J.~Kowalski-Glikman, D.~Pranzetti and G.~Rosati, ``{Symmetries of
  quantum spacetime in three dimensions},''
  \href{http://dx.doi.org/10.1103/PhysRevD.94.084044}{{\em Phys. Rev.}
  {\bfseries D94} no.~8, (2016) 084044},
\href{http://arxiv.org/abs/1606.03085}{{\ttfamily arXiv:1606.03085 [hep-th]}}.

\bibitem{GiuliaFlavio}
G.~Gubitosi and F.~Mercati, ``{Relative Locality in $\kappa$-Poincar\'e},''
  \href{http://dx.doi.org/10.1088/0264-9381/30/14/145002}{{\em Class. Quant.
  Grav.} {\bfseries 30} (2013) 145002},
\href{http://arxiv.org/abs/1106.5710}{{\ttfamily arXiv:1106.5710 [gr-qc]}}.

\bibitem{Carmona:2019fwf}
Carmona, J. M. and Cort\'es, J. L. and Relancio, J. J., ``{Relativistic
  deformed kinematics from momentum space geometry},''
  \href{http://dx.doi.org/10.1103/PhysRevD.100.104031}{{\em Phys. Rev. D}
  {\bfseries 100} no.~10, (2019) 104031},
  \href{http://arxiv.org/abs/1907.12298}{{\ttfamily arXiv:1907.12298
  [hep-th]}}.

\bibitem{Arzano:2014jua}
M.~Arzano, G.~Gubitosi, J.~Magueijo and G.~Amelino-Camelia, ``{Anti-de Sitter
  momentum space},'' \href{http://dx.doi.org/10.1103/PhysRevD.92.024028}{{\em
  Phys. Rev.} {\bfseries D92} no.~2, (2015) 024028},
\href{http://arxiv.org/abs/1412.2054}{{\ttfamily arXiv:1412.2054 [gr-qc]}}.

\bibitem{principle}
G.~Amelino-Camelia, L.~Freidel, J.~Kowalski-Glikman and L.~Smolin, ``{The
  principle of relative locality},''
  \href{http://dx.doi.org/10.1103/PhysRevD.84.084010}{{\em Phys. Rev.}
  {\bfseries D84} (2011) 084010},
\href{http://arxiv.org/abs/1101.0931}{{\ttfamily arXiv:1101.0931 [hep-th]}}.

\bibitem{Schroers2011}
B.~J.~Schroers, ``{Quantum gravity and non-commutative spacetimes in three
  dimensions: a unified approach},''
  \href{http://dx.doi.org/10.5506/APhysPolBSupp.4.379}{{\em Acta Phys. Polon.
  Supp.} {\bfseries 4} (2011) 379--402},
\href{http://arxiv.org/abs/1105.3945}{{\ttfamily arXiv:1105.3945 [gr-qc]}}.

\bibitem{Freidel:2013zga}
Freidel, Laurent and Leigh, Robert G. and Minic, Djordje, ``{Born Reciprocity
  in String Theory and the Nature of Spacetime},''
  \href{http://dx.doi.org/10.1016/j.physletb.2014.01.067}{{\em Phys. Lett. B}
  {\bfseries 730} (2014) 302--306},
  \href{http://arxiv.org/abs/1307.7080}{{\ttfamily arXiv:1307.7080 [hep-th]}}.

\bibitem{Freidel:2015pka}
Freidel, Laurent and Leigh, Robert G. and Minic, Djordje, ``{Metastring Theory
  and Modular Space-time},''
  \href{http://dx.doi.org/10.1007/JHEP06(2015)006}{{\em JHEP} {\bfseries 06}
  (2015) 006}, \href{http://arxiv.org/abs/1502.08005}{{\ttfamily
  arXiv:1502.08005 [hep-th]}}.

\bibitem{Freidel:2016pls}
Freidel, Laurent and Leigh, Robert G. and Minic, Djordje, ``{Quantum Spaces are
  Modular},'' \href{http://dx.doi.org/10.1103/PhysRevD.94.104052}{{\em Phys.
  Rev. D} {\bfseries 94} no.~10, (2016) 104052},
  \href{http://arxiv.org/abs/1606.01829}{{\ttfamily arXiv:1606.01829
  [hep-th]}}.

\bibitem{Freidel:2018apz}
Freidel, Laurent and Kowalski-Glikman, Jerzy and Leigh, Robert G. and Minic,
  Djordje, ``{Theory of metaparticles},''
  \href{http://dx.doi.org/10.1103/PhysRevD.99.066011}{{\em Phys. Rev. D}
  {\bfseries 99} no.~6, (2019) 066011},
  \href{http://arxiv.org/abs/1812.10821}{{\ttfamily arXiv:1812.10821
  [hep-th]}}.

\bibitem{Freidel:2022ryr}
Freidel, Laurent and Kowalski-Glikman, Jerzy and Leigh, Robert G. and Minic,
  Djordje, ``{Vacuum energy density and gravitational entropy},''
  \href{http://dx.doi.org/10.1103/PhysRevD.107.126016}{{\em Phys. Rev. D}
  {\bfseries 107} no.~12, (2023) 126016},
  \href{http://arxiv.org/abs/2212.00901}{{\ttfamily arXiv:2212.00901
  [hep-th]}}.

\bibitem{Berglund:2023vrm}
Berglund, Per and Geraci, Andrew and H\"ubsch, Tristan and Mattingly, David and
  Minic, Djordje, ``{Triple interference, non-linear Talbot effect and
  gravitization of the quantum},''
  \href{http://dx.doi.org/10.1088/1361-6382/ace14a}{{\em Class. Quant. Grav.}
  {\bfseries 40} no.~15, (2023) 155008},
  \href{http://arxiv.org/abs/2303.15645}{{\ttfamily arXiv:2303.15645 [gr-qc]}}.

\bibitem{Carmona:2014lqa}
Carmona, J. M. and Cort\'es, J. L. and Romeo, B., ``{Modified
  energy\textendash{}momentum conservation laws and vacuum Cherenkov
  radiation},''
  \href{http://dx.doi.org/10.1016/j.astropartphys.2015.04.009}{{\em Astropart.
  Phys.} {\bfseries 71} (2015) 21--30},
  \href{http://arxiv.org/abs/1409.8181}{{\ttfamily arXiv:1409.8181 [hep-ph]}}.

\bibitem{arzano2011anatomy}
M.~Arzano, ``{Anatomy of a deformed symmetry: Field quantization on curved
  momentum space},'' \href{http://dx.doi.org/10.1103/PhysRevD.83.025025}{{\em
  Phys. Rev.} {\bfseries D83} (2011) 025025},
\href{http://arxiv.org/abs/1009.1097}{{\ttfamily arXiv:1009.1097 [hep-th]}}.

\bibitem{lukierski1991q}
J.~Lukierski, H.~Ruegg, A.~Nowicki and V.~N.~Tolstoi, ``{q-deformation of
  Poincar\'e algebra},''
\href{http://dx.doi.org/10.1016/0370-2693(91)90358-W}{{\em Phys. Lett.}
  {\bfseries B264} (1991) 331--338}.

\bibitem{Lukierski1992344}
J.~Lukierski, A.~Nowicki and H.~Ruegg, ``{New quantum Poincar\'e algebra and
  $\kappa$-deformed field theory},''
\href{http://dx.doi.org/10.1016/0370-2693(92)90894-A}{{\em Phys. Lett.}
  {\bfseries B293} (1992) 344--352}.

\bibitem{MajidRUEGG}
S.~Majid and H.~Ruegg, ``{Bicrossproduct structure of kappa Poincare group and
  noncommutative geometry},''
  \href{http://dx.doi.org/10.1016/0370-2693(94)90699-8}{{\em Phys. Lett.}
  {\bfseries B334} (1994) 348--354},
\href{http://arxiv.org/abs/hep-th/9405107}{{\ttfamily arXiv:hep-th/9405107
  [hep-th]}}.

\bibitem{2002PhRvL..88s0403M}
J.~Magueijo and L.~Smolin, ``{Lorentz invariance with an invariant energy
  scale},'' \href{http://dx.doi.org/10.1103/PhysRevLett.88.190403}{{\em Phys.
  Rev. Lett.} {\bfseries 88} (2002) 190403},
\href{http://arxiv.org/abs/hep-th/0112090}{{\ttfamily arXiv:hep-th/0112090
  [hep-th]}}.

\bibitem{balenascience}
G.~Amelino-Camelia, G.~Gubitosi and G.~Palmisano, ``{Pathways to relativistic
  curved momentum spaces: de Sitter case study},''
  \href{http://dx.doi.org/10.1142/S0218271816500279}{{\em Int. J. Mod. Phys.}
  {\bfseries D25} no.~02, (2016) 1650027},
\href{http://arxiv.org/abs/1307.7988}{{\ttfamily arXiv:1307.7988 [gr-qc]}}.

\bibitem{2014SIGMA..10..079A}
M.~Arzano, D.~Latini and M.~Lotito, ``{Group Momentum Space and Hopf Algebra
  Symmetries of Point Particles Coupled to 2+1 Gravity},''
  \href{http://dx.doi.org/10.3842/SIGMA.2014.079}{{\em SIGMA} {\bfseries 10}
  (2014) 079},
\href{http://arxiv.org/abs/1403.3038}{{\ttfamily arXiv:1403.3038 [gr-qc]}}.

\bibitem{KowalskiGlikman:2004tz}
J.~Kowalski-Glikman and S.~Nowak, ``{Quantum $\kappa$-Poincar\'e algebra from
  de Sitter space of momenta},''
\href{http://arxiv.org/abs/hep-th/0411154}{{\ttfamily arXiv:hep-th/0411154
  [hep-th]}}.

\bibitem{tesipalmisano}
G. Palmisano, {\em Planck Scale Curved Momentum Spaces}.
\newblock PhD thesis, Universit\`a di Roma ``La Sapienza'', 2016.
\newblock
  \url{http://www.phys.uniroma1.it/fisica/sites/default/files/DOTT_FISICA/MENU/03DOTTORANDI/TesiFin28/Palmisano.pdf}.

\bibitem{2014PhRvD..90g6010B}
A.~Banburski and L.~Freidel, ``{Snyder Momentum Space in Relative Locality},''
  \href{http://dx.doi.org/10.1103/PhysRevD.90.076010}{{\em Phys. Rev.}
  {\bfseries D90} no.~7, (2014) 076010},
\href{http://arxiv.org/abs/1308.0300}{{\ttfamily arXiv:1308.0300 [gr-qc]}}.

\bibitem{Carmona:2019vsh}
Carmona, J. M. and Cort\'es, J. L. and Relancio, J. J., ``{Relativistic
  deformed kinematics from locality conditions in a generalized spacetime},''
  \href{http://dx.doi.org/10.1103/PhysRevD.101.044057}{{\em Phys. Rev. D}
  {\bfseries 101} no.~4, (2020) 044057},
  \href{http://arxiv.org/abs/1912.12885}{{\ttfamily arXiv:1912.12885
  [hep-th]}}.

\bibitem{Relancio:2022mia}
Relancio, Jos\'e Javier, ``{Relativistic deformed kinematics: From flat to
  curved spacetimes},'' \href{http://dx.doi.org/10.1142/S0219887822300045}{{\em
  Int. J. Geom. Meth. Mod. Phys.} {\bfseries 19} no.~09, (2022) 2230004},
  \href{http://arxiv.org/abs/2207.08471}{{\ttfamily arXiv:2207.08471 [gr-qc]}}.

\bibitem{Carmona:2021pxw}
Carmona, Jos\'e Manuel and Cort\'es, Jos\'e Luis and Relancio, Jos\'e Javier,
  ``{Particle\textendash{}Antiparticle Asymmetry in Relativistic Deformed
  Kinematics},'' \href{http://dx.doi.org/10.3390/sym13071266}{{\em Symmetry}
  {\bfseries 13} no.~7, (2021) 1266},
  \href{http://arxiv.org/abs/2106.05904}{{\ttfamily arXiv:2106.05904
  [hep-ph]}}.

\bibitem{Relancio:2021ahm}
Relancio, J. J., ``{Geometry of multiparticle systems with a relativistic
  deformed kinematics and the relative locality principle},''
  \href{http://dx.doi.org/10.1103/PhysRevD.104.024017}{{\em Phys. Rev. D}
  {\bfseries 104} no.~2, (2021) 024017},
  \href{http://arxiv.org/abs/2105.12573}{{\ttfamily arXiv:2105.12573 [gr-qc]}}.

\bibitem{Carmona:2021gbg}
Carmona, Jos\'e Manuel and Cort\'es, Jos\'e Luis and Relancio, Jos\'e Javier,
  ``{Curved Momentum Space, Locality, and Generalized Space-Time},''
  \href{http://dx.doi.org/10.3390/universe7040099}{{\em Universe} {\bfseries 7}
  no.~4, (2021) 99}, \href{http://arxiv.org/abs/2104.07336}{{\ttfamily
  arXiv:2104.07336 [gr-qc]}}.

\bibitem{Granik:2002wd}
A.~Granik, ``{Maguejo-Smolin transformation as a consequence of a specific
  definition of mass, velocity, and the upper limit on energy},''
\href{http://arxiv.org/abs/hep-th/0207113}{{\ttfamily arXiv:hep-th/0207113
  [hep-th]}}.

\bibitem{fock1969theory}
V. Fock, {\em The theory of space, time, and gravitation}.
\newblock Pergamon, New York, 1964.

\bibitem{2003PhRvD..67d4017M}
J.~Magueijo and L.~Smolin, ``{Generalized Lorentz invariance with an invariant
  energy scale},'' \href{http://dx.doi.org/10.1103/PhysRevD.67.044017}{{\em
  Phys. Rev.} {\bfseries D67} (2003) 044017},
\href{http://arxiv.org/abs/gr-qc/0207085}{{\ttfamily arXiv:gr-qc/0207085
  [gr-qc]}}.

\bibitem{2016IJMPS..4160126L}
I.~P.~Lobo and G.~Palmisano, ``{Geometric interpretation of
  Planck-scale-deformed co-products},''
  \href{http://dx.doi.org/10.1142/S2010194516601265}{{\em Int. J. Mod. Phys.
  Conf. Ser.} {\bfseries 41} (2016) 1660126},
\href{http://arxiv.org/abs/1612.00326}{{\ttfamily arXiv:1612.00326 [hep-th]}}.

\bibitem{moschella2006sitter}
U. Moschella, in {\em Einstein, 1905--2005: Poincar{\'e} Seminar 2005},
  pp.~120--133.
\newblock Birkh{\"a}user Basel, Basel, 2006.

\end{thebibliography}\endgroup

\end{document}